\documentclass[journal,twoside]{IEEEtran}
\usepackage{graphicx}
\usepackage{cite}
\usepackage{amsmath}
\usepackage{amssymb}
\usepackage{color}
\ifCLASSINFOpdf
\else
\fi
\newtheorem{theo}{Theorem}[section]
\newtheorem{definition}[theo]{Definition}
\newtheorem{theorem}[theo]{Theorem}
\newtheorem{lemma}[theo]{Lemma}

\begin{document}
\title{Consensus in Self-similar Hierarchical Graphs and Sierpi\'nski Graphs: Convergence Speed, Delay Robustness, and Coherence}

%\author{Michael~Shell,~\IEEEmembership{Member,~IEEE,}
%        John~Doe,~\IEEEmembership{Fellow,~OSA,}
%        and~Jane~Doe,~\IEEEmembership{Life~Fellow,~IEEE}% <-this % stops a space
%\thanks{M. Shell was with the Department
%of Electrical and Computer Engineering, Georgia Institute of Technology, Atlanta,

\author{Yi Qi, Zhongzhi Zhang, Yuhao Yi, and Huan Li% <-this % stops a space
\thanks{Manuscript received  **** **, 2017; revised **** **, 2017; accepted **** **,2017. This work was supported by the National Natural Science Foundation of China under Grant No. 61074119. This paper was recommended by Associate Editor ****.
(\emph{Corresponding author: Zhongzhi Zhang.})}
\thanks{Yi Qi, Zhongzhi Zhang, Yuhao Yi, and Huan Li are with the Shanghai Key Laboratory of Intelligent Information
Processing, School of Computer Science, Fudan University, Shanghai 200433, China (e-mail:zhangzz@fudan.edu.cn).}}


\markboth{IEEE TRANSACTIONS ON CYBERNETICS}{QI \MakeLowercase{\textit{et al.}}: Consensus in Self-similar Hierarchical Graphs and Sierpi\'nski Graphs: Convergence Speed, Delay Robustness, and Coherence}




\maketitle

\begin{abstract}
The hierarchical graphs and Sierpi\'nski graphs are constructed iteratively, which have the same number of vertices and edges at any iteration, but exhibit quite different structural properties: the hierarchical graphs are non-fractal and small-world, while the Sierpi\'nski graphs are fractal and ``large-world''. Both graphs have  found broad applications. % {\color{blue}For example, Sierpi\'nski graphs are relevant to WK-recursive networks that are widely applied in the design and implementation of parallel processing architectures.}
In this paper, we study consensus problems in hierarchical graphs and Sierpi\'nski graphs,  focusing on three important quantities of consensus problems, that is, convergence speed, delay robustness, and coherence for first-order (and second-order) dynamics, which are, respectively, determined by algebraic connectivity, maximum eigenvalue, and sum of reciprocal (and square of reciprocal) of each nonzero eigenvalue of Laplacian matrix. For both graphs, based on the explicit recursive relation of eigenvalues at two successive iterations, we evaluate the second smallest eigenvalue, as well as the largest eigenvalue, and obtain the closed-form solutions to the sum of reciprocals (and square of reciprocals) of all nonzero eigenvalues. We also compare our obtained results for consensus problems on both graphs and show that they differ in all quantities concerned, which is due to the marked difference of their topological structures.
\end{abstract}

\begin{IEEEkeywords}
Distributed average consensus, multi-agent systems, hierarchical graph, Sierpi\'nski graph, self-similar networks, graph Laplacians, convergence speed, delay robustness, network coherence.
\end{IEEEkeywords}

% \fi
\IEEEpeerreviewmaketitle

\section{Introduction}

\IEEEPARstart{A}{s}  a fundamental research object  with a long history~\cite{De74}, consensus problems cut across diverse areas of science and engineering. Typical examples include distributed computing~\cite{Ly97, CrNoGuToRi03}, load balancing~\cite{Cy89,MuGhSc98,DiFrMo99}, sensor networks~\cite{LiRu06,YuChWaYa09,ZhChLiYaGu13}, flocking~\cite{Sa06}, rendezvous~\cite{DiKy07}, vehicle formation~\cite{FaMu04} and platooning~\cite{JaLiMo03,OlMu04,Mo05}, control system technique~\cite{IlStSt14,KiAh15,DaGh15}, and synchronization of coupled oscillators~\cite{MiSt90}. In the settings of networks (graphs) of agents, consensus means that agents represented by nodes (vertices) reach agreement on a certain issue, such as pace, load, or direction and velocity. Due to their wide applications, consensus problems have received a tremendous amount of attention and made great progress in past years~\cite{SaFaMu07,WuTaCaZh16}.

In this paper, we study consensus algorithms on graphs, with emphasis on some primary aspects: convergence rate~\cite{Ol05,OlTs09,AyBa10}, delay robustness~\cite{XiWa08,MuPaAl10}, robustness to noise~\cite{XiBoKi07,BaJoMiPa08,YoScLe10,PaBa10,BaJoMiPa12,PaBa14}, which are the theme of many previous work.
All these three  issues have a close relation to the eigenvalues of Laplacian matrix for the graph on which consensus algorithms are defined.
Convergence speed measures the time of convergence of a consensus algorithm, which is closely related to the second smallest eigenvalue of Laplacian matrix~\cite{OlMu03,OlMu04}. Delay robustness refers to the ability of consensus schemes resistant to communication delay between agents, with the allowable maximum delay determined by the largest eigenvalue~\cite{OlMu03,OlMu04}. Finally, robustness to noise can be gauged by the derivation of each vertex's state from the global average of all current states, which is governed by all non-zero eigenvalues~\cite{XiBoKi07,BaJoMiPa12}.

As shown above, the three relevant issues for consensus algorithms are determined by the eigenvalues of Laplacian matrix for the underlying graphs. It is well established~\cite{GrMeSu90} that the eigenvalues of Laplacian matrix of a graph depend on its topological structure. Thus, it is of theoretical and practical interest to unveil the profound effects of structural properties of networks on their Laplacian spectrum, such as small-world feature~\cite{WaSt98} that is ubiquitous in real-life networks~\cite{Ne03}. However, it is very hard and even impossible to characterize the Laplacian spectrum of a generic graph. Most previous work about consensus problems either study part of the three problems or use the technique of  numerical simulations. Then, it makes sense to study particular graphs with ideal structure, for which the properties or behaviors of   Laplacian spectrum can be determined accurately.

In this  paper, we study  consensus problems on two families of self-similar deterministic  graphs: hierarchical graphs~\cite{BaCoDaFi09} and  Sierpi\'nski graphs~\cite{KlMi97}. The major reasons to  choose these two graphs as our research objects are as follows.  First, they have a large variety of  applications and are  extensively studied. Our research is helpful for better understanding the two important graph families.  Moreover, both graphs are constructed in an iterative way~\cite{DeKr12}, have the  same number  of nodes and the same number edges at any iteration, but exhibit strongly different
structural properties. Our work is instrumental to uncover the influence of  topological properties on several consensus schemas. Finally, the full Laplacian spectrum for both graphs can be determined iteratively, which allows to study analytically the behaviors of various  consensus algorithms that are dependent on the  eigenvalues of Laplacian matrix.  For both graphs, we provide recursive relations for the second smallest eigenvalues, the largest  eigenvalues, the sum of reciprocals (and  reciprocals of square) of all non-zero eigenvalues, based on which we further obtain and compare the asymptotic behaviors for the studied consensus algorithms, and show that the behavior difference lies on the structure  distinction of the graphs considered.



\section{Consensus problems in a graph}

Let $\mathcal{G} = (\mathcal{V},\mathcal{E})$  be an undirected connected network (graph) with $N=|\mathcal{V}|$ nodes (vertices) and $M=|\mathcal{E}|$ edges, where $\mathcal{V}$ is the  node  set and $\mathcal{E}$ is edge set. In this section, we give a brief introduction to \emph{consensus problems}~\cite{OlMu03,OlMu04} in  graph $\mathcal{G}$.

\subsection{Matrix theory}

The connectivity of a graph $\mathcal{G}$ is encoded in its adjacency matrix $\mathbf{A}$, the  entry at row $i$ ($i=1,2,\ldots,N$) and column $j$ ($j=1,2,\ldots,N$) is  $a_{ij}=1$ (or 0)  if nodes $i$ and $j$ are (not) connected by an edge. Let $N_i(\mathcal{G})$ denote the set of neighbors of node $i$. Then, the degree of node $i$ is $d_i=\sum_{j=1}^{N} a_{ij}=\sum_{j \in N_i(\mathcal{G})} a_{ij}$, and the diagonal degree matrix of $\mathcal{G}$, denoted by $\textbf{D}$, is defined as: the $i$th diagonal element is $d_i$, while all the off-diagonal elements are zeros. Thus, the Laplacian matrix $\textbf{L}$ of $\mathcal{G}$ is given by $\textbf{L}=\textbf{D}-\textbf{A}$. Let $\lambda_1$, $\lambda_2$, $\lambda_3$, $\cdots$, $\lambda_N$ be the $N$ eigenvalues of matrix $\textbf{L}$ rearranged in an increasing order, that is  $\lambda_1<\lambda_2\leq \cdots \leq \lambda_N$. Since $\mathcal{G}$ is connected, its Laplacian matrix $\textbf{L}$ always has a zero eigenvalue with a single degeneracy, i.e., $\lambda_1= 0$, and all other eigenvalues are larger than zero, among which $\lambda_2$ is named the algebraic connectivity of the graph by Fiedler~\cite{Fi7303}, thus also called Fiedler eigenvalue.

\subsection{Consensus problems}

We next introduce several linear consensus problems under different assumptions.

\subsubsection{Consensus without delay and noise}

Let us consider graph $\mathcal{G}$ as a multi-agent system, where an agent and available information flow between two agents are, respectively, looked upon as a node and an edge in $\mathcal{G}$. We assume that the system of dynamic agents is described by $\dot{x}_i(t)=u_i(t)$, where $x_i(t)\in \mathbb{R}$ and $u_i(t) \in \mathbb{R}$ denote, respectively, the state of agent $i$ and the associated control input. It was shown by Olfati-Saber and Murray~\cite{OlMu03,OlMu04} that the following linear dynamic system
\begin{equation}\label{Con01}
\dot{x}_i(t)=\sum_{j \in N_i(\mathcal{G})}\left[x_j(t)-x_i(t)\right]
\end{equation}
with a collective dynamics
\begin{equation}\label{Con02}
\dot{x} (t)=-\textbf{L}x (t) \,
\end{equation}
solves a consensus problem. In other words, the state of all agents in system~({\ref{Con02}}) asymptotically converges to the average value  $\bar{x}=(1/N )\sum_{i=1}^N x_i(0)$, where $x_i(0)$ is the initial state of node $i$.

The convergence speed of system~({\ref{Con01}}) can be measured by the algebraic connectivity $\lambda_2$:  the larger the value of $\lambda_2$,  the faster the  convergence speed.

\subsubsection{Consensus under communication time-delay}

As in many real-world situations, the communication between pairs of agents is often not instantaneous~\cite{HuKoSz10}. Instead, an agent reacts to the information or signal received from its neighbors with some finite time lag. Suppose that agent $i$ receives a message sent by one of its neighbors $j$ after a time delay $\tau_{ij}$. Then, the dynamics of the system is governed by~\cite{OlMu03,OlMu04}
\begin{equation}\label{Con03}
\dot{x}_i(t)=\sum_{j \in N_i(\mathcal{G})}\left[x_j(t-\tau_{ij})-x_i(t-\tau_{ij})\right]\,.
\end{equation}
Throughout this  paper, we focus on the case of uniform delay, where the time delay $\tau_{ij}$ for all pairs of nodes $i$ and $j$ is fixed to $\tau$. Then, system~(\ref{Con03}) becomes
\begin{equation}\label{Con04}
\dot{x}_i(t)=\sum_{j \in N_i(\mathcal{G})}\left[x_j(t-\tau)-x_i(t-\tau)\right]\,
\end{equation}
with a collective dynamics
\begin{equation}\label{Con05}
\dot{x} (t)=-\textbf{L}x (t-\tau) \,.
\end{equation}
Olfati-Saber and Murray~\cite{OlMu04} have proven that in the system~(\ref{Con05}) with an undirected and connected network topology, all nodes globally asymptotically reach an average consensus if and only if the following condition is satisfied:
\begin{equation}\label{Con06}
0\leq \tau< \tau_{\rm max}=\frac{\pi}{2\lambda_n} \,.
\end{equation}

Equation~(\ref{Con06}) shows that the analysis of consensus problem in a connected undirected network with  an equal time-delay in all links is reduced to spectral analysis of Laplacian matrix of the network. Specifically, the largest eigenvalue $\lambda_n$ is a measure of delay robustness for achieving a agreement in a network: the smaller the largest eigenvalue  $\lambda_n$, the bigger the maximum delay $\tau_{\rm max}$, and vice versa. On the other hand, similarly to system~({\ref{Con01}}),   the convergence speed of system~({\ref{Con05}}) is also determined by $\lambda_2$, which converges fast with the increasing of $\lambda_2$.

\subsubsection{Consensus with white noise}

Since autonomous systems must operate in uncertain environments without direct supervision, it is important that such systems should be robust with respect to environment uncertainty and communication uncertainty. Thus, it is of great interest to consider how robust distributed consensus algorithms are to external disturbances. For both first-order and second-order systems, robustness to uncertainty and noise can be quantified using the quantity called \emph{network coherence} in terms of $\mathcal{H}_2$ norm~\cite{XiBoKi07,BaJoMiPa08,YoScLe10,PaBa11,BaJoMiPa12}.

\emph{First-order noisy consensus}. In the first-order consensus problem, each node has a single state subject to stochastic disturbances (noise). For simplicity, we assume that every agent is independently affected by white  noise of the same intensity. The resulting system is an extension of system~({\ref{Con02}}) given by
\begin{equation}\label{Con07}
\dot{x} (t)=-\textbf{L}x (t)+w(t)\,,
\end{equation}
where $w(t) \in \mathbb{R}^N$ is a random signal with zero-mean and unit variance.
In contrast to standard consensus problem without noise, instead of converging to the average of the initial state values, the sequence of node states $x(t)$ becomes a stochastic process and fluctuates around the average of the current node states. The variance of
these fluctuations can be captured by network coherence. Without loss of generality, we consider the case $\sum^N_{i=1}x_i(0)=0$.

\begin{definition}
For a graph $\mathcal{G}$, the \emph{first-order network coherence} $H_{1}(\mathcal{G})$ is defined as the mean,
steady-state variance of the deviation from the average of all node
values,
\[
H_{1}(\mathcal{G}):=\frac{1}{N}\sum^N_{i=1}\lim_{t\to\infty}{\bf{{E}}}
\left\{x_i(t)-\frac{1}{N}\sum^N_{j=1}x_j(t)\right\}.
\]
\end{definition}

%Let $\bf{\Pi}$ be the projection matrix onto the subspace of $\mathbb{R}^N$ orthogonal to the vector with order $N$ of all ones, $\textbf{1}_N=[1,1,\ldots,1]^{\top} \in \mathbb{R}^N$. Thus $\bf{\Pi}=\textbf{I}_N-\frac{1}{N}{\textbf{1}_N \textbf{1}_N^{\top}}$, where $\textbf{I}_N$ is the $N\times N$ identity matrix. We define the output of the system (\ref{Con07}) to be
%\begin{equation}\label{Con08}
%y(t)=\bf{\Pi} x(t).
%\end{equation}
%Then, $H^{(1)}$ is related to the $H_2$ norm of the system described by
%(\ref{Con07}) and (\ref{Con08}) as follows,
%\begin{equation}\label{Con09}
%H^{(1)}=\frac{1}{N}{\bf
%tr}\left(\int_0^{\infty}e^{-L^*t}\Pi e^{-Lt}dt\right)\,.
%\end{equation}

It has been established~\cite{XiBoKi07,BaJoMiPa08,YoScLe10,PaBa11,BaJoMiPa12} that $H_{1}(\mathcal{G})$ is fully determined by the $N-1$ non-zero eigenvalues for Laplacian matrix $\textbf{L}$. Specifically, the network coherence of the
first-order system is given by
\begin{equation}\label{Con10}
H_{1}(\mathcal{G})=\frac{1}{2N}\sum^{N}_{i=2}\frac{1}{\lambda_i}\,.
\end{equation}
Lower $H_{1}(\mathcal{G})$ implies good robustness of the system irrespective of the presence of noise, that is, nodes remain closer to consensus at the average of their current states. %We note that in addition to the first-order consensus, the value of $H^{(1)}$ also has important means in many other fields such as electrical networks~\cite{DoSn84}, molecular networks~\cite{GuMo96}, and random walks~\cite{ChRaRuSmTi97}.

In addition to the coherence of  first-order noisy consensus,  the sum $\Lambda_{\rm sum}=\sum^{N}_{i=2}\frac{1}{\lambda_i}$ of the $N-1$ non-zero eigenvalues for Laplacian matrix $\textbf{L}$ of a graph $\mathcal{G}$ also plays a key role in determining many other interesting properties of the graph. For example, as a  measure of overall connectedness and robustness~\cite{DeMeRoSaVa18}  of a graph, the Kirchhoff index~\cite{KlRa93,LiLiCa16}  defined as the sum of resistance distances between all the $N(N-1)/2$ pairs of vertices, is equal to $N\Lambda_{\rm sum}$~\cite{Te91,ChRaRuSm89}. Again for instance, as a metric of  diffusion velocity or mean cost of search in a graph, the average of hitting times for random walks  over all the $N(N-1)$ pairs of vertices equals  $\frac{2M\Lambda_{\rm sum}}{N-1}$~\cite{ChRaRuSm89}, which can thus be expressed in terms of  $H_{1}(\mathcal{G})$ as $\frac{4MN}{N-1}H_{1}(\mathcal{G})$.

\emph{Second-order noisy consensus}.  In the second-order consensus problem, each node $i$ has two states, position $x_i(t)$
and velocity $v_i(t)$. For example, in the vehicular formation problem, each vehicle has a position and a velocity.  The objective is for each vehicle to
move with constant heading velocity while keeping a fixed,
pre-specified distance between itself and all of its neighbors. In a second-order system, the
node states  consist of a position vector $x(t)$ and a velocity vector $v(t)$.  The states are measured relative to the heading velocity $\bar{v}$ and position $\bar{x}(t)$. The equation governing the system dynamics is given by
\begin{equation}\label{Con11}
\left[ \begin{array}{c}
\dot{x}(t)\\
\dot{v}(t)
\end{array} \right]=\left[ \begin{array}{cc}
\textbf{0} & \textbf{I}_n\\
-\textbf{L} & -\textbf{L}
\end{array} \right]\left[ \begin{array}{c}
x(t)\\
v(t)
\end{array} \right]+\left[ \begin{array}{c}
\textbf{0}\\
\textbf{I}_n
\end{array} \right]w(t),
\end{equation}
where $w(t) \in \mathbb{R}^{N}$ is a mutually uncorrelated white noise process. Notice that in system~(\ref{Con11}) stochastic perturbations enter only in the velocity terms.

The network coherence of the second-order system is defined in terms
of the node positions only, and it captures how closely the
vehicle formation follows the desired heading trajectory in steady-state.
\begin{definition}
For a graph $\mathcal{G}$, the \emph{second-order network coherence} $H_{2}(\mathcal{G})$ is the mean, steady-state
variance of the deviation of each vehicle's position error from the
average of all vehicle position errors.
\end{definition}

%Let the output for the system~(\ref{Con11}) be
%\begin{equation}\label{Con12}
%y(t)=\left[ \begin{array}{cc} \bf{\Pi} & \textbf{0}
%\end{array} \right]\left[ \begin{array}{c}
%x(t)\\
%v(t)
%\end{array} \right].
%\end{equation}
As in the case of first-order dynamics, the variance can be related to the $\mathcal{H}_2$ norm of
the system defined by~\eqref{Con11}, and its value also depends on the nonzero eigenvalues of the Laplacian
matrix $\mathbf{L}$, given by~\cite{BaJoMiPa08,PaBa11,BaJoMiPa12}
\begin{equation}\label{Con13}
H_{2}(\mathcal{G})=\frac{1}{2N}\sum^{N}_{i=2}\frac{1}{\lambda_i^2}.
\end{equation}
A small $H_{2}(\mathcal{G})$ corresponds to good robust second-order system of agents subject to stochastic disturbances.

Thus, the interesting quantities for consensus problems such as convergence speed, delay robustness, coherence for both first-order and second-order dynamics, are all dependent on the eigenvalues of the Laplacian matrix. %, which are related to system structures and system size.

\subsection{Related work}

We first review some related work on convergence rate and robustness of communication delay for consensus algorithms. Olfati-Saber demonstrated that it is possible to dramatically increase the second largest eigenvalue of the Laplacian matrix of a regular graph by rewiring some edges, without significantly decreasing the  largest eigenvalue~\cite{Ol05}.
Olshevsky and Tsitsiklis  provided lower bounds on the worst-case convergence time for
many different types of linear, time-invariant, distributed consensus algorithms~\cite{OlTs09}. M\"{u}nz~\textit{et al.} studied the robustness of consensus schemes with different feedback
delays and proposed a scalable delay-dependent design algorithm for consensus controllers for a  class of linear
multi-agent systems~\cite{MuPaAl10}.

For the first-order noisy consensus problem, using the measure in~(\ref{Con10}) Young~\textit{et al.} derived analytical expressions for the coherence in path, cycle, and star graphs~\cite{YoScLe10}. Patterson~\textit{et al.} provided a series of theoretical study on the coherence in fractal trees (T-fractal and Vicsec fractal)~\cite{PaBa11,PaBa14}, and tori and lattices of different dimensions~\cite{BaJoMiPa12}, with the mean square performance and robust yet fragile nature of torus also being considered by Ma and Elia~\cite{MaEl15}. Very recently, we presented analytical solutions for network coherence in the small-world Farey graph~\cite{YiZhLiCh15,ZhCo11}, as well as the scale-free small-world Koch network~\cite{YiZhShCh17}. Summers~\textit{et al.} addressed the graph topology design problem~\cite{SuShLyDo15}: For a given graph, choose a fixed number of edges added to it, with an object  to minimize the coherence of the resultant graph.

In contrast to the first-order dynamics, related work about second-order consensus problem is relatively less. Patterson and Bamieh studied the network coherence for two fractal trees---T-fractal and Vicsec fractal~\cite{PaBa11,PaBa14}, and also considered the coherence for tori and lattices with their collaborators~\cite{BaJoMiPa12}. We addressed the same problem in Koch network and compared the result with that in regular ring lattice with  identical average degree 3~\cite{YiZhShCh17}.

Previous work showed that the interested quantities (convergence rate, maximum communication delay, and the coherence of both first-order and second-order noisy consensus problems) are  closely related to the structural properties of  underlying networks, and unveiled partial influences of some particular features on the behavior of consensus on networks, such as small-world phenomenon~\cite{Ol05,YiZhLiCh15}, fractal dimension~\cite{PaBa11,PaBa14}. However, the diversity of realistic network leads to the existence of diverse structural properties. For example, some real networks (e.g. power grid~\cite{CaJaSo01}) have an exponentially-decaying degree distribution~\cite{AmScBaSt00}, although many real-life networks have a power-law degree distribution~\cite{BaAl99}. To the best our knowledge, extensive analytical research about various consensus algorithms in small-world graphs with an exponential degree distribution is much less. The consensus problems in such graphs are not well characterized.

Motivated by previous work, in the sequel, we will study consensus problems in two much studied self-similar networks:  hierarchical graphs~\cite{BaCoDaFi09} and  Sierpi\'nski graphs~\cite{KlMi97}.  Both graphs have   the same  number  of nodes and  edges  but exhibit  different structures. For example, hierarchical graphs have a degree distribution of an exponential form, while the degree of all vertices in Sierpi\'nski graphs is identical except some special vertices.  Moreover, the eigenvalues of Laplacian matrices for both graphs can be determined analytically, which permits us to  determine explicitly relevant quantities for consensus problems, to show how they scale with the system size, and further to explore the impacts of network architectures  on the consensus problems.

\section{Network construction and properties}

In this section, we introduce two families of self-similar graphs that are generated iteratively by different assembly mechanisms~\cite{DeKr12,BaCoDaFi09,KlMi97}.
The first class of graphs is hierarchical graphs~\cite{BaCoDaFi09} generated by a branching iteration; and  the
second type of graphs  is Sierpi\'nski graphs~\cite{KlMi97} created by a nested iteration. Both networks have the
same numbers of nodes and edges, but display quite different
structural properties. Thus, they are good candidate networks
for studying consensus problems, in order to unveil the effects
of topologies on consensus performance.

%The first network is  hierarchical graphs $H(n,k)$, which is based on a branching iteration. And the second network is Sierpi\'nski graphs $S(n,k)$, which is based on a nested iteration. Both networks have the same numbers of nodes and edges, but display quite different structural properties. Thus, they are good candidate networks for us to study the consensus problems on them and unveil the effects of topologies on consensus performance.

\subsection{Hierarchical graphs}

The hierarchical graphs are created by the hierarchical product of graphs introduced first by Godsil and McKay~\cite{GoMc78}.
\begin{definition}~\cite{GoMc78}
 \label{def1} Let $\mathcal{G}_1(\mathcal{V}_1,\mathcal{E}_1)$ and $\mathcal{G}_2(\mathcal{V}_2,\mathcal{E}_2)$ be two graphs, with a root vertex labeled by $1$. Then, the hierarchical product $\mathcal{G}_2 \sqcap \mathcal{G}_1$ of $\mathcal{G}_1$ and $\mathcal{G}_2$ is a graph with vertices $x_2x_1$, $x_i \in \mathcal{V}_i$ $(i=1,2)$, and edges $(x_2x_1, y_2y_1)$, where either $y_2=x_2$ and $y_1$ and $ x_1$ are adjacent  in $\mathcal{G}_1$ or $y_1=x_1=1$ and $y_2$ and $ x_2$ are adjacent in $\mathcal{G}_2$.
\end{definition}

Note that the hierarchical product is associative~\cite{BaCoDaFi09},  we can iteratively apply the operation of hierarchical product to any graph. The hierarchical graphs are those graphs generated by iteratively using the operation of hierarchical product to the complete graph.

\begin{definition}\label{def2a}~\cite{BaCoDaFi09}
Let $\mathcal{K}_k$ denote the complete graph with $k$ ($k \geq 3$) vertices, labelled by 1, 2, $\ldots$, $k$. Then the hierarchical graphs $\mathcal{H}(n,k)$, $n\geq 1$, are the hierarchical product of $n$ replicas of $\mathcal{K}_k$, with  $\mathcal{H}(1,k)$ isomorphic to $\mathcal{K}_k$. That is, $\mathcal{H}(n,k)=\mathcal{H}(n-1,k)  \sqcap \mathcal{K}_k=\mathcal{K}_k\sqcap\stackrel{n}{\cdots}\sqcap \mathcal{K}_k$.
\end{definition}

\begin{figure}
\centering
\includegraphics[width=0.85\linewidth,trim=0 0 0 0]{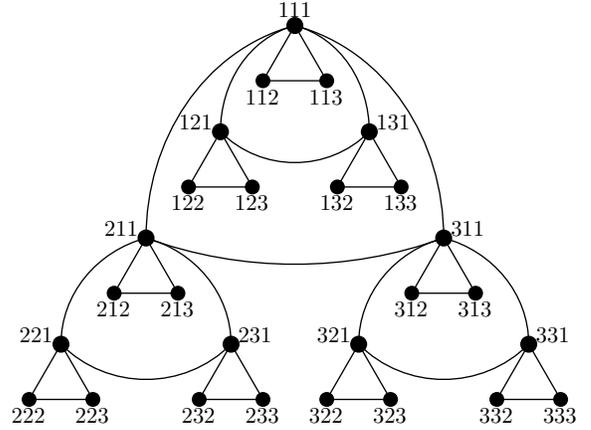}
\caption{Illustration of a hierarchical graph $\mathcal{H}(3,3)$.}\label{Hiraconstuct}
\end{figure}

Fig.~\ref{Hiraconstuct} illustrates a hierarchical graph $\mathcal{H}(3,3)$.

%%%%%%%%%%%%%%%%%%%%%%%%%%%%%%%%%%%%%%%%%%%%%%%%%%%%%
%%%%%%%%%%%%%%%%%%%%%%%%%%%%%%%%%%%%%%%%%%%%%%%%%%%%%

%In addition to applying hierarchical product,
The hierarchical graphs $\mathcal{H}(n,k)$ can be alternatively constructed in the following way~\cite{liDoQiZh15}.

\begin{definition}\label{def2b}
For $n=1$,  $\mathcal{H}(1,k)$ is  the complete graph $\mathcal{K}_k$.  For $n> 1$, $\mathcal{H}(n,k)$ is obtained from $\mathcal{H}(n-1,k)$:  for each vertex in $\mathcal{H}(n-1,k)$, we create a complete graph $\mathcal{K}_{k-1}$ and connect all its $k-1$ vertices to their mother  vertex in  $\mathcal{H}(n-1,k)$.
\end{definition}

%Let $B_n$ denote the hierarchical graphs after $n$ iterations. The process begins with a complete graph $K_k$ corresponding to the generation $n=1$. For $n> 1$, $B_n$ is obtained from $B_{n-1}$ by creating a complete graph $K_{k-1}$ for each vertex in $B_{n-1}$ and connecting it to the $k-1$ vertices of the complete graph $K_{k-1}$.

Informally, a graph is self-similar, if it displays identical structure at every scale. For a
formal definition of self-similar graph, we refer the reader to~\cite{Kr05}. It has been shown that many real complex networks are self-similar~\cite{SoHaMa05}.  The hierarchical graphs have a self-similar structure. In $\mathcal{H}(n,k)$, there are $k$ vertices with highest degree  $n(k-1)$.  We called them hub vertices.  Given $\mathcal{H}(n-1,k)$, $\mathcal{H}(n,k)$  can be obtained by performing the following operations. First, we create a complete graph $\mathcal{K}_k$, and generate $k$ copies of $\mathcal{H}(n-1,k)$.  %,which are labeled as $\mathcal{H}(n-1,k)^{(1)}$, $\mathcal{H}(n-1,k)^{(2)}$, $\ldots$, $\mathcal{H}(n-1,k)^{(k)}$, respectively.
Then, we identify the $k$ vertices of the complete graph $\mathcal{K}_k$ and  $k$ hub vertices of  $k$ different replicas of $\mathcal{H}(n-1,k)$. %Fig.~\ref{Hireconstruct2} illustrates the amalgamation process.

%%%%%%%%%%%%%%%%%%%%%%%%%%%%%%%%%%%%%%%%%%%%%%%%%%%%%
%\begin{figure}
%\centering
%\includegraphics[width=0.9\linewidth,trim=0 0 0 0]{hiercons2.eps}
%\caption{Another construction approach of  hierarchical graphs highlighting their self-similarity.}\label{Hireconstruct2}
%\end{figure}
%%%%%%%%%%%%%%%%%%%%%%%%%%%%%%%%%%%%%%%%%%%%%%%%%%%%%

Let $N_{n}$ and $E_{n}$ denote the order (number of vertices) and size (number of edges) of hierarchical graphs $\mathcal{H}(n,k)$. According to the construction algorithms, we have
\begin{equation}\label{hierconstruct1}
N_n=k^n,
\end{equation}
and
\begin{equation}\label{hierconstruct5}
E_n=\frac{k^{n+1}-k}{2},
\end{equation}

In the hierarchical graphs $\mathcal{H}(n,k)$, the degree spectrum
is discrete:  the number $N(\delta,n)$ of vertices of
degree  $n(k-1)$, $(n-1)(k-1)$, $(n-2)(k-1)$, $\ldots$, $2(k-1)$, $k-1$,   is
$k$, $(k-1)k$, $(k-1)k^{2}$, $\ldots$, $(k-1)k^{n-2}$, $(k-1)k^{n-1}$, respectively. Thus, the degree of vertices in the hierarchical graphs $\mathcal{H}(n,k)$ follows an exponential distribution, with its cumulative  degree distribution~\cite{Ne03} $P_{{\rm cum}}(\delta) = \sum_{\delta' \geq \delta}N(\delta',n)/ N_n \sim k^{1-\frac{\delta}{k-1}}$ decaying exponentially with the degree $\delta$ but independent of $n$. Such a degree distribution has been previously observed in some real technology networks~\cite{AmScBaSt00,CaJaSo01}.
Besides, the hierarchical graphs $\mathcal{H}(n,k)$ exhibit the typical small-world characteristics of real-life networks~\cite{WaSt98}. They have high average clustering coefficient, and both their diameter and average distance grow logarithmically with the network order~\cite{liDoQiZh15}.
%Hence, the hierarchical graph is a good model mimicking real networks.

The hierarchical graphs have been used to mimic  real networks, such as biological networks~\cite{DeKr12} and polymer networks~\cite{liDoQiZh15}.

\subsection{Sierpi\'nski graphs}

The Sierpi\'nski graphs $\mathcal{S}(n,k)=(\mathcal{V}(\mathcal{S}(n,k)),\mathcal{E}(\mathcal{S}(n,k)))$ ($n \ge 1$ and $k \ge 3$) were introduced by Klav\v{z}ar and Milutinovi\'c~\cite{KlMi97} as a two-parametric generalization of the Tower of Hanoi graph~\cite{ZhWuLiCo16,JiLiZh17}. They are defined on the vertex set comprising of all $n$-tuples of integers $1,2,\cdots,k$, that is, $\mathcal{V}(\mathcal{S}(n,k)) = \{1,2,\cdots,k\}^n$. All  vertices in  $\mathcal{S}(n,k)$ can be  labelled in the form  $u_1u_2\cdots u_n$, where $u_i \in \{1,2,\cdots,k\}$ for all $i=1,2,\cdots,n$. Two vertices   $p=p_1p_2\cdots p_n$ and $q=q_1q_2\cdots q_n$ are connected to each other by an edge if and only if there exists an integer $h$ $(1\le h \le n)$ such that

(a) $p_i = q_i$ for $1 \le i \le h-1$;

(b) $p_h \ne q_h$;

(c) $p_i = q_h$ and $p_i = q_h$ for $h+1 \le i \le n$.

\begin{figure}
\begin{center}
\includegraphics[width=3.0in]{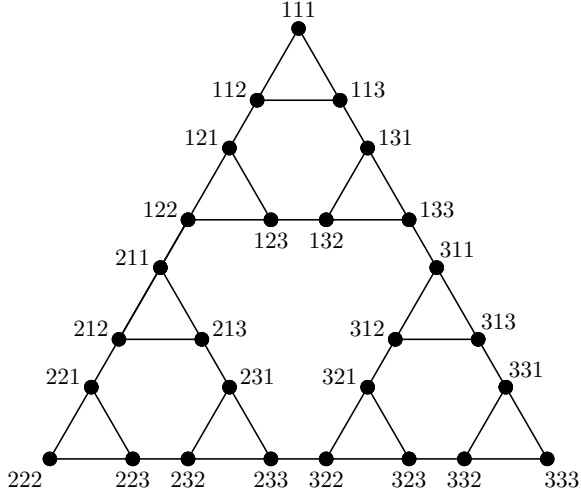}
\end{center}
\caption[kurzform]{The Sierpi\'nski graph $\mathcal{S}(3,3)$ and its vertex labeling.}\label{sierconstruction}
\end{figure}

Fig.~\ref{sierconstruction} illustrates the Sierpi\'nski graph $\mathcal{S}(3,3)$.

%%%%%%%%%%%%%%%%%%%%%%%%%%%%%%%%%%%%%%%%%%%%%%%%%%%%%%%%%%
%%%%%%%%%%%%%%%%%%%%%%%%%%%%%%%%%%%%%%%%%%%%%%%%%%%%%%%%%%

%It is obvious that the Sierpi\'nski graphs $\mathcal{S}(n,k)$ has a self-similar structure, which can be constructed in another method. In order to obtain $\mathcal{S}(n,k)$, we can amalgamate $k$ copies of $\mathcal{S}(n-1,k)$ by adding extra edges connecting the corresponding extreme vertices with the degree $k-1$ in the replicas of each $\mathcal{S}(n-1,k)$.

The order and size of the Sierpi\'nski graphs $\mathcal{S}(n,k)$ are identical to those corresponding to hierarchical graphs $\mathcal{H}(n,k)$. In other words, $N_n=k^n$ and $E_n=\frac{k^{n+1}-k}{2}$.
%\begin{equation}\label{sierconstruct1}
%N_n=k^n,
%\end{equation}
%and
%\begin{equation}\label{sierconstruct5}
%E_n=\frac{k^{n+1}-k}{2}\,.
%\end{equation}

Analogously to  hierarchical graphs  $\mathcal{H}(n,k)$,  Sierpi\'nski graphs $\mathcal{S}(n,k)$ are also self-similar. However, Sierpi\'nski graphs $\mathcal{S}(n,k)$ have different structural properties from those of hierarchical graphs $\mathcal{H}(n,k)$. First, Sierpi\'nski graphs are more homogeneous. In $\mathcal{S}(n,k)$, there are $k$ vertices of label form $i i \cdots  i$ $(1 \le i \le k)$, which  are  called extreme vertices. Each extreme vertex has a degree $k-1$,  while the degree of any other  vertex  is $k$.
Second, Sierpi\'nski graphs are not small-world. The diameter~\cite{KlMi97} of  $\mathcal{S}(n,k)$ is $2^n-1$ increasing exponentially with $n$, and thus growing as a power-law function of  network order $N_n$.

%An important notion for a self-similar graph is the \emph{fractal dimension},  defined as follows~\cite{Te90}.
\begin{definition}~\cite{Te90}
Let $\mathcal{G} = (\mathcal{V},\mathcal{E})$  be a connected undirected graph.
For any vertex $v \in \mathcal{V}$,  define $B(r)$ as the  radius $r$, centered at $v$, i.e.,
$B(r) = \{u \in \mathcal{V}: d(u,v) \leq r\}$, where $d(u,v)$ denotes the shortest-path length  between the two vertices $u$ and $v$ in $\mathcal{G}$. The \emph{fractal dimension} of $\mathcal{G}$ is
\[
d_f(\mathcal{G}) := - \limsup_{r \rightarrow \infty} \frac{\log B(r) }{\log r}.
\]
\end{definition}

Sierpi\'nski graphs $\mathcal{S}(n,k)$ are fractal~\cite{MaPe97,KlMo05}, with their fractal dimension being $d_f (\mathcal{S}(n,k))=\frac{\log{k}}{\log{2}}$.
%\begin{equation}\label{sierconstruct10}
%d_f (\mathcal{S}(n,k))=\frac{\log{k}}{\log{2}}\,. \nonumber
%\end{equation}
%and
%\begin{equation}\label{sierconstruct11}
%d_s = \frac{2\log{k}}{\log{(k+2)}},
%\end{equation}
%respectively~\cite{RaTo83}.

Except for the intrinsic theoretical interest~\cite{HiKlMiPeSt13,HiklSa17}, Sierpi\'nski graphs  play an important role in topology, mathematics, and computer science.   For example, adding an ``open edge'' to each of the $k$ extreme vertices in Sierpi\'nski graphs yields the WK-recursive networks. These resultant networks have many favorable structure properties, including a high degree of regularity, symmetry and scalability, which can be applied as a model of interconnection networks widely used in implementing large-scale distributed systems~\cite{FaImSa08,WaJu15}. Recently, great efforts have been dedicated to investigating various issues of the WK-Recursive networks, such as topological properties~\cite{FaWaHu07}, broadcasting algorithms~\cite{FaLiChNg07}, and fault tolerance~\cite{Fu05}.

\section{Consensus algorithms in  hierarchical graphs}

In this section, we study analytically several quantities  related to consensus algorithms in  hierarchical graphs $\mathcal{H}(n,k)$, concentrating  on convergence speed of first-order consensus without delay and noise, delay robustness of first-order consensus, as well as coherence of first-order and second-order noisy consensus algorithms. For all these consensus algorithms, we derive explicit or recursive expressions for those quantities concerned, as well as their asymptotic scalings.

\subsection{Spectra of Laplacian matrix}

Let $\textbf{L}_n$ denote the Laplacian matrix of  hierarchical graphs $\mathcal{H}(n,k)$. And let $\Psi_n$ represent the set of eigenvalues of $\textbf{L}_n$  defined  as
\begin{equation}\label{hiersp2}
\Psi_n=\left\{\lambda_{1}^{(n)}, \lambda_{2}^{(n)}, \cdots, \lambda_{N_n}^{(n)}\right\},
\end{equation}
satisfying $0=\lambda_{1}^{(n)}\le \lambda_{2}^{(n)}\le \cdots\le \lambda_{N_n}^{(n)}$.
Note that $\Psi_n$ is a multiset. That is, the distinctness of   elements in $\Psi_n$ is neglected. It has been shown~\cite{liDoQiZh15} that all eigenvalues of $\textbf{L}_n$  can be determined recursively.

%With an aim to determine the first-order and second-order coherence in the hierarchical graphs $B_n$, we introduce the Laplacian spectra~\cite{liDoQiZh15} of $B_n$ as  follows:

\begin{lemma}\label{hiereig}
All the nonzero eigenvalues in the  set $\Psi_n$ of matrix $\textbf{L}_n$ can be classified into two subsets $\Psi_n^{(1)}$ and $\Psi_n^{(2)}$, satisfying $\Psi_n=\Psi_n^{(1)} \cup \Psi_n^{(2)} \cup \{0\}$, where $\Psi_n^{(1)}$ contains only eigenvalue $k$ with multiplicity $(k-2)k^{n-1}+1$,
$\Psi_n^{(2)}$ includes $2k^{n-1}-2$ elements $\widetilde{\lambda}_{i}^{(n)}$ and $\widetilde{\lambda}_{i'}^{(n)}$ generated by $\lambda_{i}^{(n-1)}$, $i=2,3,\ldots,k^{n-1}$, through the following relations
\begin{align}\label{hiereig1}
\widetilde{\lambda}_{i}^{(n)}=&\frac{1}{2}\bigg(-\sqrt{k^2+2k\lambda_{i}^{(n-1)}+\left(\lambda_{i}^{(n-1)}\right)^2-4\lambda_{i}^{(n-1)}}+\nonumber\\
&k+\lambda_{i}^{(n-1)}\bigg),
\end{align}
\begin{align}\label{hiereig5}
\widetilde{\lambda}_{i'}^{(n)}=&\frac{1}{2}\bigg(\sqrt{k^2+2k\lambda_{i}^{(n-1)}+\left(\lambda_{i}^{(n-1)}\right)^2-4\lambda_{i}^{(n-1)}}+\nonumber\\
&k+\lambda_{i}^{(n-1)}\bigg),
\end{align}
which obey  relation $\widetilde{\lambda}_{2}^{(n)}\le\widetilde{\lambda}_{3}^{(n)}\le\cdots\le\widetilde{\lambda}_{k^{n-1}}^{(n)}<k-2<\widetilde{\lambda}_{2'}^{(n)}\le\widetilde{\lambda}_{3'}^{(n)}\le\cdots\le\widetilde{\lambda}_{{k^{(n-1)}}'}^{(n)}$.
\end{lemma}
For $\mathcal{H}(1,k)$, the set of eigenvalues of its Laplacian matrix is
\begin{equation}\label{am28}
\Psi_1=\{0,\underbrace{k,k,k,\dots,k}_{k-1\mbox{}}\}.
\end{equation}
Then, by iteratively applying Lemma~\ref{hiereig}, we can obtain all the $k^n$ eigenvalues of the Laplacian matrix for the hierarchical graphs $\mathcal{H}(n,k)$ for all $n \geq 1$.

\subsection{Convergence speed and delay robustness}

Let $\epsilon_n$ and $\zeta_n$ be, respectively,  the second smallest eigenvalue and the largest eigenvalue for Laplacian matrix of hierarchical graphs $\mathcal{H}(n,k)$.  As shown above,   $\epsilon_n$ and $\zeta_n$ measure  the convergence speed and delay robustness of first-order consensus algorithms on  hierarchical graphs $\mathcal{H}(n,k)$. In this subsection, we characterize the  asymptotic behaviors  of the  two critical eigenvalues $\epsilon_n$ and $\zeta_n$.

\begin{theorem}\label{Hierdelay}
For hierarchical graphs $\mathcal{H}(n,k)$, when $n$ is sufficient large, the asymptotic behaviors of the second smallest eigenvalue $\epsilon_n$ and the largest  eigenvalue $\zeta_n$,  are
\begin{equation}\label{hierdelay2}
\epsilon_n \approx k^{2-n}
\end{equation}
and
\begin{equation}\label{hierdelay1}
\zeta_n \approx (k-1)n,
\end{equation}
respectively.
\end{theorem}

\begin{IEEEproof}
By~\eqref{hiereig1},   the second smallest eigenvalue $\epsilon_n$ of the Laplacian matrix for  hierarchical graphs $\mathcal{H}(n,k)$ obeys  recursion relation
\begin{align}\label{hierdelay5}
\epsilon_n&=\frac{1}{2}\left(k+\epsilon_{n-1}-\sqrt{k^2+2k\epsilon_{n-1}+\epsilon_{n-1}^2-4\epsilon_{n-1}}\right)\nonumber\\
&=\frac{1}{2}\left(k+\epsilon_{n-1}-k\sqrt{1+\frac{2\epsilon_{n-1}}{k}+\frac{\epsilon_{n-1}^2-4\epsilon_{n-1}}{k^2}}\right).
\end{align}
Analogously,  the largest eigenvalue $\zeta_n$ satisfies
\begin{equation}\label{hierdelay3}
\zeta_{n}=\frac{1}{2}\left(\zeta_{n-1} \sqrt{1+\frac{k^2}{\zeta_{n-1}^2}+\frac{2k-4}{\zeta_{n-1}}}+k+\zeta_{n-1}\right).
\end{equation}
Applying the following  relation
\begin{equation}\label{hierdelay70}
\sqrt{1-x}=1-\frac{1}{2}x+o(x),\quad  (x \to 0),
\end{equation}
to~\eqref{hierdelay5} and~\eqref{hierdelay3} and ignoring the higher order infinitesimal in the radical terms lead to
\begin{equation}\label{hierdelay11}
\epsilon_n \approx \frac{\epsilon_{n-1}}{k}
\end{equation}
and
\begin{equation}\label{hierdelay10}
\zeta_{n} \approx \zeta_{n-1}+k-1.
\end{equation}
Considering the fact that $\epsilon_{1}=\zeta_{1}=k$, we obtain the asymptotic behaviors of $\epsilon_n$ and $\zeta_{n}$  given in~\eqref{hierdelay2} and ~\eqref{hierdelay1}. % $n \to \infty$:
%\begin{equation}\label{hierdelay13}
%\epsilon_n \approx k^{2-n}
%\end{equation}
%and
%\begin{equation}\label{hierdelay12}
%\zeta_n \approx (k-1)n,
%\end{equation}
%respectively.
%This completes the proof.
\end{IEEEproof}

\begin{figure}[htbp]
\centering
\begin{minipage}{4.3cm}
\centering
\includegraphics[width=1.0\linewidth,trim=0 0 0 0]{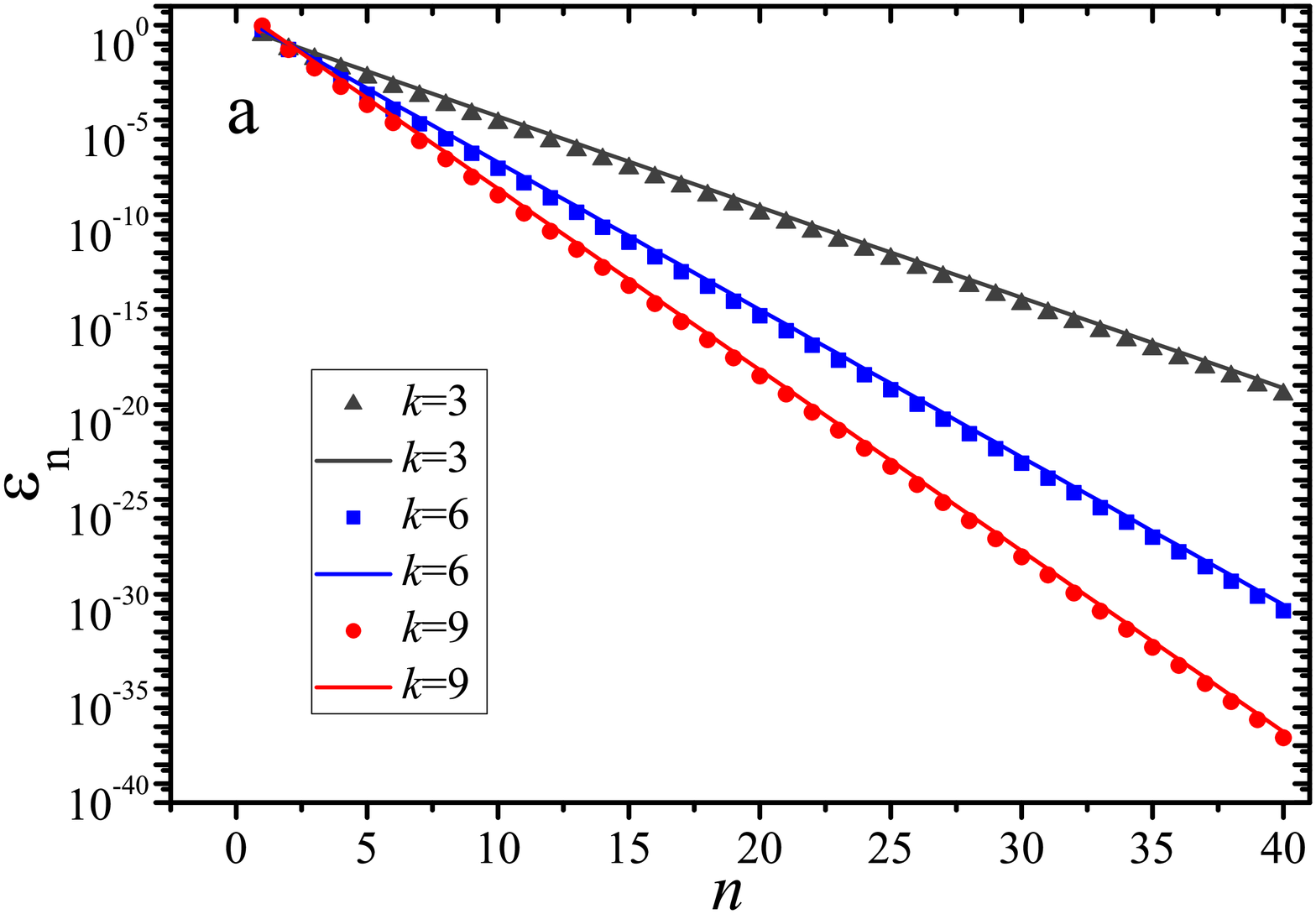}
\end{minipage}
\begin{minipage}{4.3cm}
\centering
\includegraphics[width=1.0\linewidth,trim=0 0 0 0]{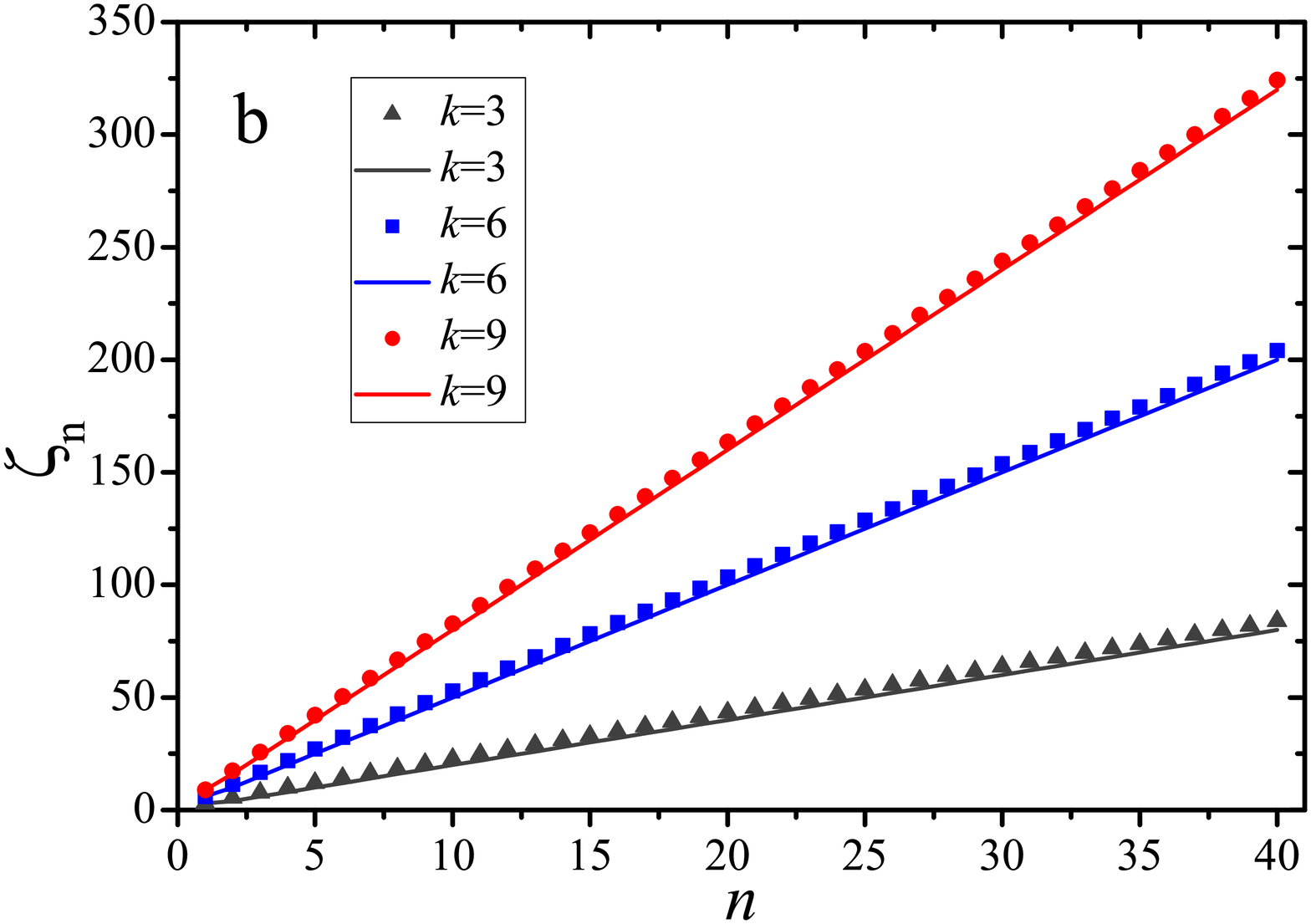}
\end{minipage}
\caption[kurzform]{The second smallest eigenvalue $\epsilon_n$ (a) and the largest eigenvalues $\zeta_n$ (b) for hierarchical graphs $\mathcal{H}(n,k)$ with various $n$ and $k$.}\label{hierminmax}
\end{figure}

%In Figs.~\ref{hiermin} and~\ref{hiermax},
In Fig.~\ref{hierminmax},  we report the exact values and  approximative results for the second smallest eigenvalue $\epsilon_n$  and the largest eigenvalue  $\zeta_n$  corresponding to hierarchical graphs $\mathcal{H}(n,k)$ with various $n$ and $k$.
In the figure, solid symbols represent the exact results iteratively generated by~\eqref{hierdelay5} or~\eqref{hierdelay3},  while the straight lines denote the approximative results given by~\eqref{hierdelay2} or~\eqref{hierdelay1}.
It can be seen that both the exact and  approximate  results agree well with each other.  Moreover, as is expected, the approximate values of the largest eigenvalue (the second smallest eigenvalue) are   slightly  smaller (larger) than the corresponding exact ones, since we ignore the infinitesimal of higher order during the derivation of the approximate formulas.

%Merge into one figure below.
%%%%%%%%%%%%%%%%%%%%%%%%%%%%%%%%%%%%%%%%%%%%%%%%%%%%%%
%\begin{figure}
%\centering
%\includegraphics[width=1.0\linewidth,trim=0 0 0 0]{hiermin.eps}
%\caption{The second smallest eigenvalue $\epsilon_n$  for hierarchical graphs $\mathcal{H}(n,k)$ with various $n$ and $k$.}\label{hiermin}
%\end{figure}
%%%%%%%%%%%%%%%%%%%%%%%%%%%%%%%%%%%%%%%%%%%%%%%%%%%%%%
%
%%%%%%%%%%%%%%%%%%%%%%%%%%%%%%%%%%%%%%%%%%%%%%%%%%%%%%
%\begin{figure}
%\centering
%\includegraphics[width=1.0\linewidth,trim=0 0 0 0]{hiermax.eps}
%\caption{The largest eigenvalues $\zeta_n$  for hierarchical graphs $\mathcal{H}(n,k)$ with various $n$ and $k$.}\label{hiermax}
%\end{figure}
%%%%%%%%%%%%%%%%%%%%%%%%%%%%%%%%%%%%%%%%%%%%%%%%%%%%%%

%%%%%%%%%%%%%%%%%%%%%%%%%%%%%%%%%%%%%%%%%%%%%%%%%%%%%
%%%%%%%%%%%%%%%%%%%%%%%%%%%%%%%%%%%%%%%%%%%%%%%%%%%%%

Theorem~\ref{Hierdelay} shows that,  as $n\rightarrow \infty$, the second smallest eigenvalue $\epsilon_n$ is  inversely proportional to the order $N_n$, satisfying $\epsilon_n \approx  k^2N_n^{-1}$,  while the largest eigenvalue $\zeta_n$ increases logarithmically with $N_n$.

\subsection{First-order and second-order coherence}

We proceed to determine the explicit expressions and their leading behaviors for the coherence of the first-order and second-order noisy consensus algorithms in  hierarchical graphs $\mathcal{H}(n,k)$, which are denoted as $H_1(\mathcal{H}(n,k))$ and $H_2(\mathcal{H}(n,k))$, respectively.

\subsubsection{First-order coherence}

With the known results about the eigenvalues for Laplacian matrix of hierarchical graphs $\mathcal{H}(n,k)$, we can obtain the first-order network coherence.
\begin{theorem}\label{Hiercoh1order}
For the hierarchical graphs $\mathcal{H}(n,k)$ with $N_n$ vertices, the first-order coherence of the system with dynamics defined in~\eqref{Con07} is
\begin{equation}\label{Hiercoh01}
H_1(\mathcal{H}(n,k))=\frac{(2n-1)k^n-2nk^{n-1}+1}{2k^{n+1}}\,.
\end{equation}
In the limit of  large $n$, $H_1(\mathcal{H}(n,k))$ grows with network order $N_n$ as
\begin{equation}\label{Hiercoh02}
\lim_{n \to \infty}H_1(\mathcal{H}(n,k)) = \frac{k-1}{k^2} \log_k{N_n}.
\end{equation}
\end{theorem}

\begin{IEEEproof}
Based on the previously established result, the first-order coherence of $\mathcal{H}(n,k)$ is
\begin{equation}\label{Hiercoh1}
H_1(\mathcal{H}(n,k))=\frac{1}{2N_n}\sum^{N_n}_{i=2}\frac{1}{\lambda_i^{(n)}},
\end{equation}
where $\lambda_i^{(n)}$, $2\le i\le N_n$, are all the non-zero eigenvalues
of the Laplacian matrix for  $\mathcal{H}(n,k)$. We now determine the sum on the right-hand side (rhs) of~\eqref{Hiercoh1}, denoted as $\Lambda_n$, which can be evaluated as
%\begin{equation}\label{Hiercoh5}
%\Lambda_n=\sum^{V_n}_{i=2}\frac{1}{\lambda_i^{(n)}},
%\end{equation}
%According to the previous result, we can rewrite ~\eqref{Hiercoh5} as
\begin{equation}\label{Hiercoh11}
\Lambda_n=\sum^{N_n}_{i=2}\frac{1}{\lambda_i^{(n)}}=\sum_{\lambda_i^{(n)} \in \Psi_n^{(1)}}\frac{1}{\lambda_i^{(n)}}+
\sum_{\widetilde{\lambda}_{i}^{(n)} \in \Psi_n^{(2)}}\frac{1}{\widetilde{\lambda}_{i}^{(n)}}.
\end{equation}
Let $\Lambda_n^{(1)}$ and $\Lambda_n^{(2)}$ stand for the two sum terms on the rhs of~\eqref{Hiercoh11}.
For $\lambda_i^{(n)} \in \Psi_n^{(1)}$, we have
\begin{equation}\label{Hiercoh15}
\Lambda_n^{(1)}=\sum_{\lambda_i^{(n)} \in \Psi_n^{(1)}}\frac{1}{\lambda_i^{(n)}}=(k-2)k^{n-2}+\frac{1}{k}.
\end{equation}

For  $\widetilde{\lambda}_{i}^{(n)} \in \Psi_n^{(2)}$, it can be evaluated as follows. From Lemma~\ref{hiereig}, each eigenvalue $\lambda_i^{(n-1)}$ in $\Psi_{n-1}$ gives rise to two eigenvalues $\widetilde{\lambda}_{i}^{(n)}$ and $\widetilde{\lambda}_{i'}^{(n)}$ in $\Psi_n^{(2)}$, which obey relations $\widetilde{\lambda}_{i}^{(n)}+\widetilde{\lambda}_{i'}^{(n)}=k+\lambda_i^{(n-1)}$ and $\widetilde{\lambda}_{i}^{(n)} \cdot \widetilde{\lambda}_{i'}^{(n)}=\lambda_i^{(n-1)}$.
%\begin{equation}\label{Hiercoh21}
%\widetilde{\lambda}_{i}^{(n)}+\widetilde{\lambda}_{i'}^{(n)}=k+\lambda_i^{(n-1)}
%\end{equation}
%and
%\begin{equation}\label{Hiercoh25}
%\widetilde{\lambda}_{i}^{(n)} \cdot \widetilde{\lambda}_{i'}^{(n)}=\lambda_i^{(n-1)}.
%\end{equation}
Then,
\begin{equation}\label{Hiercoh31}
\frac{1}{\widetilde{\lambda}_{i}^{(n)}}+\frac{1}{\widetilde{\lambda}_{i'}^{(n)}}
=\frac{\widetilde{\lambda}_{i}^{(n)}+\widetilde{\lambda}_{i'}^{(n)}}{\widetilde{\lambda}_{i}^{(n)}\cdot\widetilde{\lambda}_{i'}^{(n)}}
=\frac{k}{\lambda_i^{(n-1)}}+1.
\end{equation}
%It should be pointed that the eigenvalue $\lambda_{1}^{(n-1)}=0$ in $\Psi_{n-1}$ gives rise two eigenvalues i.e. $0$ and $k$ in $\Psi_n^{(2)}$.
Consequently, we have
\begin{equation}\label{Hiercoh35}
\Lambda_n^{(2)}=\sum_{\widetilde{\lambda}_{i}^{(n)} \in \Psi_n^{(2)}}\frac{1}{\widetilde{\lambda}_{i}^{(n)}}=k\cdot\Lambda_{n-1}+N_{n-1}-1.
\end{equation}
Combining~\eqref{Hiercoh11},~\eqref{Hiercoh15} and~\eqref{Hiercoh35}, we obtain the  recursive relation for $\Lambda_n$ as
$\Lambda_n=k\Lambda_{n-1}+2(k-1)k^{n-2}-\frac{k-1}{k}$,
%\begin{equation}\label{Hiercoh40}
%\Lambda_n=k\Lambda_{n-1}+2(k-1)k^{n-2}-\frac{k-1}{k}.
%\end{equation}
which, together  the initial condition $\Lambda_1=\frac{k-1}{k}$, is solved to yield
\begin{equation}\label{Hiercoh41}
\Lambda_n=\frac{(2n-1)k^n-2nk^{n-1}+1}{k}.
\end{equation}
Substituting~\eqref{Hiercoh41} into~\eqref{Hiercoh1}, we obtain the explicit expression for the first-order coherence in the hierarchical graphs $\mathcal{H}(n,k)$:
\begin{equation}\label{Hiercoh45}
H_1(\mathcal{H}(n,k))=\frac{(2n-1)k^n-2nk^{n-1}+1}{2k^{n+1}}\,.
\end{equation}
For large $n$,~\eqref{Hiercoh45} implies
\begin{equation}\label{Hiercoh51}
\lim_{n \to \infty}H_1(\mathcal{H}(n,k))=\frac{k-1}{k^2}n.
\end{equation}
Because $n=\log_k{N_n}$, for $n \to \infty$, $H_1(\mathcal{H}(n,k))$ can be expressed in terms of  $N_n$ as
\begin{equation}\label{Hiercoh55}
\lim_{n \to \infty}H_1(\mathcal{H}(n,k)) = \frac{k-1}{k^2} \log_k{N_n}.
\end{equation}
This completes the proof.
\end{IEEEproof}
Theorem~\ref{Hiercoh1order} indicates that for large $n$, the first-order coherence of hierarchical graphs $\mathcal{H}(n,k)$ grows as a logarithmic function of $N_n$.

%{\color{red}
%By using the relationship between the $H_1(\mathcal{G})$ and $T(\mathcal{G})$ defined in the ~\eqref{Con104}, we can obtain the mean hitting time of hierarchical graphs $\mathcal{H}(n,k)$ as follows.
%\begin{theorem}\label{Hiercohrw}
%For the hierarchical graphs $\mathcal{H}(n,k)$ with $N_n$ vertices, the mean hitting time $T(\mathcal{H}(n,k))$ scales with the number of vertices as
%\begin{equation}\label{Hiercohrw1}
%T(\mathcal{H}(n,k))\thicksim \frac{2(k-1)}{k}N_n\log_k{N_n}\,.
%\end{equation}
%\end{theorem}
%}

\subsubsection{Second-order coherence}

We now derive the second-order coherence in the hierarchical graphs $\mathcal{H}(n,k)$.
\begin{theorem}\label{Hiercoh2order}
For the hierarchical graphs $\mathcal{H}(n,k)$, the second-order coherence of the system with dynamics defined in~\eqref{Con11} is
\begin{align}\label{Hiercoh2a}
H_2(\mathcal{H}(n,k))=&\frac{k-k^2+k^n(k^2-5k-6)+k^{2n}(4k+6)}{2k^{n+3}(1+k)}\nonumber\\
& +\frac{2n(1-k)}{k^3}.
\end{align}
In the limit of large $n$, the leading term of $H_2(\mathcal{H}(n,k))$ can be represented in terms of  network order $N_n$ as
\begin{equation}\label{Hiercoh2b}
H_2(\mathcal{H}(n,k)) \sim  \frac{2k+3}{k^3(k+1)}N_n.
\end{equation}
\end{theorem}

\begin{IEEEproof}
By definition,
\begin{equation}\label{Hiercoh61}
H_2(\mathcal{H}(n,k))=\frac{1}{2N_n}\sum^{N_n}_{i=2}\frac{1}{\left(\lambda_i^{(n)}\right)^2}.
\end{equation}
Let $\Gamma_n$ denote the sum term on the rhs of~\eqref{Hiercoh61}. Then,  $\Gamma_n$  can be evaluated as
\begin{equation}\label{Hiercoh65}
\Gamma_n=\sum_{\lambda_i^{(n)} \in \Psi_n^{(1)}}\frac{1}{\left(\lambda_i^{(n)}\right)^2}+
\sum_{\widetilde{\lambda}_{i}^{(n)} \in \Psi_n^{(2)}}\frac{1}{\left(\widetilde{\lambda}_{i}^{(n)}\right)^2}.
\end{equation}
We denote the two sum terms on the rhs of~\eqref{Hiercoh65}  by $\Gamma_n^{(1)}$ and $\Gamma_n^{(2)}$, respectively.  $\Gamma_n^{(1)}$ can be expressed as
\begin{equation}\label{Hiercoh71}
\Gamma_n^{(1)}=\sum_{\lambda_i^{(n)} \in \Psi_n^{(1)}}\frac{1}{\left(\lambda_i^{(n)}\right)^2}=(k-2)k^{n-3}+\frac{1}{k^2}.
\end{equation}
Considering %from~\eqref{Hiercoh21} and~\eqref{Hiercoh25}, we have
\begin{equation}\label{Hiercoh75}
\frac{1}{\left(\widetilde{\lambda}_{i}^{(n)}\right)^2}+\frac{1}{\left(\widetilde{\lambda}_{i'}^{(n)}\right)^2}
%&=\frac{\left(\widetilde{\lambda}_{i}^{(n)}+\widetilde{\lambda}_{i'}^{(n)}\right)^2-2\widetilde{\lambda}_{i}^{(n)}\cdot\widetilde{\lambda}_{i'}^{(n)}}{\left(\widetilde{\lambda}_{i}^{(n)}\cdot\widetilde{\lambda}_{i'}^{(n)}\right)^2}\nonumber\\
=\frac{(\lambda_{i}^{(n-1)}+k)^2-2\lambda_{i}^{(n-1)}}{(\lambda_{i}^{(n-1)})^2},
\end{equation}
$\Gamma_n^{(2)}$ can be evaluated as
\begin{equation}\label{Hiercoh81}
\Gamma_n^{(2)}
%&=\sum_{\lambda_{i}^{(n-1)} \in \Psi_{n-1}\setminus\{0\}}\frac{(\lambda_{i}^{(n-1)}+k)^2-2\lambda_{i}^{(n-1)}}{\left(\lambda_{i}^{(n-1)}\right)^2}\nonumber\\
=\sum_{\lambda_{i}^{(n-1)} \in \Psi_{n-1}\setminus\{0\}}\left(1+\frac{k^2}{\left(\lambda_{i}^{(n-1)}\right)^2}+\frac{2k-2}{\lambda_{i}^{(n-1)}} \right).
\end{equation}
Plugging~\eqref{Hiercoh71} and~\eqref{Hiercoh81} into~\eqref{Hiercoh65} gives
\begin{equation}\label{Hiercoh85}
\Gamma_n=k^2\Gamma_{n-1}+(2k-2)\Gamma_{n-1}+k^{n-1}-1,
\end{equation}
which under the initial condition $\Gamma_1=\frac{k-1}{k^2}$ is solved inductively to yield
\begin{align}\label{Hiercoh91}
\Gamma_n=&\frac{k-k^2+k^n(k^2-5k-6)+k^{2n}(4k+6)}{k^3(k+1)}\nonumber\\
&+4n\cdot k^{n-3}(1-k).
\end{align}
Inserting the expression for $\Gamma_n$ into $H_2(\mathcal{H}(n,k))=\frac{\Gamma_n}{2N_n}$ gives~\eqref{Hiercoh2a}.
%Then, the closed-form solution to the second-order coherence in the hierarchical graphs $\mathcal{H}(n,k)$ is
%\begin{align}\label{Hiercoh95}
%H_2(\mathcal{H}(n,k))=&\frac{k-k^2+k^n(k^2-5k-6)+k^{2n}(4k+6)}{2k^{n+3}(1+k)}\nonumber\\
%&+\frac{2n(1-k)}{k^3}.
%\end{align}
When $n \to \infty$,  the dominating term of  $H_2(\mathcal{H}(n,k))$ increases with  $N_n$ as
\begin{equation}\label{Hiercoh101}
H_2(\mathcal{H}(n,k)) \sim \frac{2k+3}{k^3(k+1)}N_n.
\end{equation}
This completes the proof.
\end{IEEEproof}

Theorem~\ref{Hiercoh2order}  means that when the network order $N_n$ is sufficiently  large, the second-order coherence $H_2(\mathcal{H}(n,k))$ behaves linearly with $N_n$.

\section{Consensus algorithms in Sierpi\'nski graphs}

%In the proceeding section,  we  studied  consensus algorithms in the hierarchical graphs, including convergence speed of first-order consensus without delay and noise, delay robustness of first-order consensus, as well as coherence of first-order and second-order noisy consensus algorithms.
In this section, we study  consensus algorithms in  Sierpi\'nski graphs. We are concerned with the same quantities as those corresponding  to the  hierarchical graphs. We will show that the behaviors of related quantities are significantly different from those associated with  the  hierarchical graphs.

\subsection{Spectra of Laplacian matrix}

In the case without confusion, we use the same notations as those  corresponding to the hierarchical graphs. Let $\textbf{L}_n$ denote the Laplacian matrix of the Sierpi\'nski graphs $\mathcal{S}(n,k)$, the eigenvalue  set of which is dented by $\Phi_n$, given by
\begin{equation}\label{siersp2}
\Phi_n=\left\{\lambda_{1}^{(n)}, \lambda_{2}^{(n)}, \cdots, \lambda_{N_n}^{(n)}\right\},
\end{equation}
satisfying $0=\lambda_{1}^{(n)}\le \lambda_{2}^{(n)}\le \cdots\le \lambda_{N_n}^{(n)}$.

The eigenvalues of the Laplacian matrix for  Sierpi\'nski graphs $\mathcal{S}(n,k)$ have been fully determined~\cite{MaPe97}, which have been applied to numerous aspects,  e.g., relaxation dynamics~\cite{BlJu02}. In~\cite{MaPe97},  a  recursive relation for the eigenvalues of Laplacian matrix for  Sierpi\'nski graphs $\mathcal{S}(n,k)$ was provided, as stated in the following lemma.

\begin{lemma}\label{Siereig}
For Sierpi\'nski graphs $\mathcal{S}(n,k)$, $n\ge 2$, the non-zero eigenvalues in  set $\Phi_{n}$ can be classified into two subsets $\Phi_{n}^{(1)}$ and $\Phi_{n}^{(2)}$, satisfying $\Phi_{n}=\Phi_{n}^{(1)} \cup \Phi_{n}^{(2)} \cup \{0\}$, where $\Phi_{n}^{(1)}$ consists of two eigenvalues $k$ and $k+2$, the multiplicities of which are
$\frac{k-2}{2}\left(k^{n-1}+\frac{k}{k-2}\right)$ and
$\frac{k-2}{2}\left(k^{n-1}-1 \right)$,
respectively; while $\Phi_{n}^{(2)}$ contains the rest $2k^{n-1}-2$ nonzero eigenvalues. Moreover, each eigenvalue $\lambda_i^{(n-1)}$,  $2\le i \le k^{n-1}$, in  set $\Phi_{n-1}$ generates two elements in set $\Phi_{n}^{(2)}$,  both of which have the same degeneracy as that of $\lambda_i^{(n-1)}$ and are the roots of  the following equation in $\lambda$:
\begin{equation}\label{Siercoh3}
\lambda^2-(k+2)\lambda+\lambda_{i}^{(n-1)}=0.
\end{equation}
\end{lemma}

Since  $\mathcal{S}(1,k)$ is isomorphic to $\mathcal{H}(1,k)$, the set of eigenvalues for  Laplacian matrix of $\mathcal{S}(1,k)$ is
\begin{equation}\label{am280}
\Phi_1=\Psi_1=\{0,\underbrace{k,k,k,\dots,k}_{k-1\mbox{}}\}. \nonumber
\end{equation}
Then, by recursively using Lemma~\ref{Siereig}, we obtain all the eigenvalues for $\mathcal{S}(n,k)$ for any $n>1$.

\subsection{Convergence speed and delay robustness}

For the Laplacian matrix of Sierpi\'nski graphs $\mathcal{S}(n,k)$, the second smallest eigenvalue $\epsilon_n$ and the largest eigenvalue $\zeta_{n}$ can be analytically determined as given in the following theorem.
\begin{theorem}\label{sierdelay}
For Sierpi\'nski graphs $\mathcal{S}(n,k)$, the largest eigenvalue of its Laplacian matrix $\mathbf{L}_n$ is $\zeta_{n}=k+2$ for all $n \geq 2$, while the second smallest eigenvalue $\epsilon_n$ of $\mathbf{L}_n$ satisfies
\begin{equation}\label{sierdelay1}
\epsilon_{n} \approx \frac{k}{(k+2)^{n-1}},
\end{equation}
as $n \to \infty$.
\end{theorem}

\begin{IEEEproof}
Let $f_1(x)$ and $f_2(x)$ be two functions of real number $x$ in interval  $[0,k+2]$, defined by
\begin{equation}\label{sierdelay31}
f_1(x)=\frac{k+2-\sqrt{(k+2)^2-4x}}{2}
\end{equation}
and
\begin{equation}\label{sierdelay32}
f_2(x)=\frac{k+2+\sqrt{(k+2)^2-4x}}{2},
\end{equation}
respectively.  It is easy to see that function $f_1(x)$ (or $f_2(x)$) is a monotonically  increasing (or decreasing) function on its domain $x \in [0,k+2]$, satisfying $0\le f_1(x)\le \frac{k+2}{2} \le f_2(x)\le k+2$.

According to Lemma~\ref{Siereig}, each eigenvalue $\lambda_{i}^{(n-1)}$ in set $\Phi_{n-1}$ gives rise to two eigenvalues $\widetilde{\lambda}_{i}^{(n)}$ and $\widetilde{\lambda}_{i'}^{(n)}$ in set $\Phi_{n}^{(2)}$ through~\eqref{Siercoh3},
with $\widetilde{\lambda}_{i}^{(n)}=f_1(\lambda_{i}^{(n-1)})$ and $\widetilde{\lambda}_{i'}^{(n)}=f_2(\lambda_{i}^{(n-1)})$.
For any eigenvalue $\widetilde{\lambda}_{i}^{(n)} \in \Phi_n^{(2)}$, we have
\begin{equation}\label{sierdelay21}
\widetilde{\lambda}_{i}^{(n)}=\frac{1}{2}\left(k+2+\sqrt{(k+2)^2-4\lambda_{i}^{(n-1)}}\right).
\end{equation}
Since all the  eigenvalues of a connected graph are greater than or equal to zero,~\eqref{sierdelay21} means that all  the eigenvalues of Laplacian matrix for Sierpi\'nski graphs $\mathcal{S}(n,k)$ is less than or equals
$k+2$. Thus,  the largest eigenvalue  of  Laplacian matrix for $\mathcal{S}(n,k)$  is always $\zeta_n=k+2$ for $n \geq 2$.

For the second smallest eigenvalue $\epsilon_{n}$, it obeys the following  recursive relation:
\begin{equation}\label{sierdelay2}
\epsilon_{n}=\frac{1}{2}\left(k+2-(k+2)\sqrt{1-\frac{4\epsilon_{n-1}}{(k+2)^2}}\right).
\end{equation}
Using the approximate relation in~\eqref{hierdelay70} and applying a similar argument for the proof of Theorem~\ref{Hierdelay}, we obtain
\begin{equation}\label{sierdelay3}
\epsilon_{n} \approx \frac{\epsilon_{n-1}}{k+2}.
\end{equation}
Considering  $\epsilon_{1}=k$,~\eqref{sierdelay3} leads to~\eqref{sierdelay1}.
%\begin{equation}\label{sierdelay4}
%\epsilon_{n} \approx \frac{k}{(k+2)^{n-1}}.
%\end{equation}
%This completes the proof.
\end{IEEEproof}

%%%%%%%%%%%%%%%%%%%%%%%%%%%%%%%%%%%%%%%%%%%%%%%%%%%%%
\begin{figure}
\centering
\includegraphics[width=1.0\linewidth,trim=0 0 0 0]{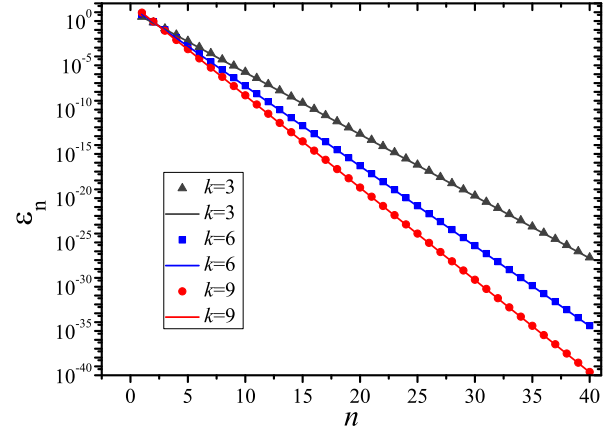}
\caption{The second smallest eigenvalue $\epsilon_{n}$ of the Laplacian matrix for Sierpi\'nski graphs $\mathcal{S}(n,k)$ with various $n$ and $k$.}\label{Siermin}
\end{figure}
%%%%%%%%%%%%%%%%%%%%%%%%%%%%%%%%%%%%%%%%%%%%%%%%%%%%%

In Fig.~\ref{Siermin}, we report the comparison of approximate and exact results for the second smallest eigenvalue $\epsilon_{n}$ of the Laplacian matrix for Sierpi\'nski graphs $\mathcal{S}(n,k)$, which are generated by~\eqref{sierdelay1} and~\eqref{sierdelay2}, respectively. In Fig.~\ref{Siermin}, solid symbols denote the accurate results, while the straights lines represent the approximate ones. Fig.~\ref{Siermin} shows that the results yielded by~\eqref{sierdelay1} and~\eqref{sierdelay2} are consistent with each other, the difference between which is intangible.

Theorem~\ref{sierdelay} implies that as $n\rightarrow \infty$, the algebraic connectivity $\epsilon_n$ of Sierpi\'nski graphs $\mathcal{S}(n,k)$ is a power-law function of $N_n$ as $\epsilon_n \approx k(k+2)(N_n)^{-\log_k(k+2)}$, which is much smaller than the algebraic connectivity $k^2N_n^{-1}$ corresponding to hierarchical graphs $\mathcal{H}(n,k)$. Thus, the speed of convergence of the consensus algorithm described by~(\ref{Con05}) in hierarchical graphs is considerably faster than in Sierpi\'nski graphs.

On the other hand, the largest eigenvalue $\zeta_n$ for the Laplacian matrix of $\mathcal{S}(n,k)$ is a constant $k+2$, independent of $n$. While for hierarchical graphs $\mathcal{H}(n,k)$,
$\zeta_n$ grows linearly with $n$ as $\zeta_n \approx (k-1)n$, much larger than $k+2$. Therefore,  the consensus algorithm described by~\eqref{Con05} in Sierpi\'nski graphs $\mathcal{S}(n,k)$ is more robust to delay than in hierarchical graphs $\mathcal{H}(n,k)$.

\subsection{First-order and second-order coherence}

Let $H_1(\mathcal{S}(n,k))$ and $H_2(\mathcal{S}(n,k))$ denote, respectively, the coherence of the first-order and second-order noisy consensus algorithms in  Sierpi\'nski graphs $\mathcal{S}(n,k)$. Below we determine their accurate expressions and leading scalings.

\subsubsection{First-order coherence}

Using the above recursive relations related to the eigenvalues for Laplacian matrix of Sierpi\'nski graphs $\mathcal{S}(n,k)$ at two successive iterations, we can deduce the first-order coherence.
\begin{theorem}\label{siercoh1order}
For the Sierpi\'nski graphs $\mathcal{S}(n,k)$ with $N_n$ vertices, the first-order coherence of the system with dynamics defined in~\eqref{Con07} is
\begin{align}\label{Siercoh4a}
H_1(\mathcal{S}(n,k))=&\frac{(k^2+k+2)(k-1)(k+2)^n-4k}{4k^{n+1}(k+1)(k+2)}-\nonumber\\
&\frac{(k-2)(k+1)}{4k(k+2)}.
\end{align}
In the limit of  large $n$, $H_1(\mathcal{S}(n,k))$ scales with the network order $N_n$ as
\begin{equation}\label{Siercoh4b}
\lim_{n \to \infty}H_1(\mathcal{S}(n,k))= \frac{k^3+k-2}{4k(k+1)(k+2)}N_n^{\frac{\log (k+2)}{\log k}-1}.
\end{equation}
\end{theorem}

\begin{IEEEproof}
Let $\Theta_n$ denote the sum of reciprocals of all the $N_n-1$ nonzero eigenvalues of Laplacian matrix for $\mathcal{S}(n,k)$. Then, $H_1(\mathcal{S}(n,k))=\frac{\Theta_n}{2N_n}$. By definition,
\begin{equation}\label{Siercoh5}
\Theta_n=\sum_{i=2}^{N_n}\frac{1}{\lambda_i^{(n)}}=\sum_{\lambda_i^{(n)} \in \Phi_n^{(1)}}\frac{1}{\lambda_i^{(n)}}+\sum_{\widetilde{\lambda}_{i}^{(n)} \in \Phi_n^{(2)}}\frac{1}{\widetilde{\lambda}_{i}^{(n)}}.
\end{equation}
Denote the two sum terms on the rhs of~\eqref{Siercoh5} as $\Theta_n^{(1)}$ and $\Theta_n^{(2)}$, respectively. By Lemma~\ref{Siereig}, the first sum is
\begin{equation}\label{Siercoh6}
\Theta_n^{(1)}=\frac{k-2}{2}\left(k^{n-2}+\frac{1}{k-2}+\frac{k^{n-1}-1}{k+2}\right).
\end{equation}

We continue to compute the second sum term $\Theta_n^{(2)}$. According to  Vieta's formulas, the two roots  $\widetilde{\lambda}_{i,1}^{(n)}$ and $\widetilde{\lambda}_{i,2}^{(n)}$ of~\eqref{Siercoh3} obey relations $\widetilde{\lambda}_{i,1}^{(n)}+\widetilde{\lambda}_{i,2}^{(n)}=k+2$ and $\widetilde{\lambda}_{i,1}^{(n)}\cdot\widetilde{\lambda}_{i,2}^{(n)}=\lambda_{i}^{(n-1)}$,
%\begin{equation}\label{Siercoh7}
%\widetilde{\lambda}_{i,1}^{(n)}+\widetilde{\lambda}_{i,2}^{(n)}=k+2
%\end{equation}
%and
%\begin{equation}\label{Siercoh8}
%\widetilde{\lambda}_{i,1}^{(n)}\cdot\widetilde{\lambda}_{i,2}^{(n)}=\lambda_{i}^{(n-1)},
%\end{equation}
which indicate
\begin{equation}\label{Siercoh9}
\frac{1}{\widetilde{\lambda}_{i,1}^{(n)}}+\frac{1}{\widetilde{\lambda}_{i,2}^{(n)}}
=\frac{\widetilde{\lambda}_{i,1}^{(n)}+\widetilde{\lambda}_{i,2}^{(n)}}{\widetilde{\lambda}_{i,1}^{(n)}\cdot\widetilde{\lambda}_{i,2}^{(n)}}
=\frac{k+2}{\lambda_{i}^{(n-1)}}.
\end{equation}
Then the second sum term $\Theta_n^{(2)}$ in~\eqref{Siercoh5} can be evaluated as
\begin{equation}\label{Siercoh10}
\Theta_n^{(2)}=\sum_{\widetilde{\lambda}_{i}^{(n)} \in \Phi_n^{(2)}}\frac{1}{\widetilde{\lambda}_{i}^{(n)}}
=\sum_{\lambda_{i}^{(n-1)} \in \Phi_{n-1}\setminus\{0\}}\frac{k+2}{\lambda_{i}^{(n-1)}}.
\end{equation}
Inserting~\eqref{Siercoh6} and~\eqref{Siercoh10} into~\eqref{Siercoh5} yields
\begin{equation}\label{Siercoh11}
\Theta_n=(k+2)\Theta_{n-1}+\frac{k-2}{2}\left(k^{n-2}+\frac{1}{k-2}+\frac{k^{n-1}-1}{k+2}\right).
\end{equation}
Using the initial condition $\Theta_1=\frac{k-1}{k}$,~\eqref{Siercoh11} is solved to obtain
\begin{align}\label{Siercoh12}
\Theta_n=&\frac{(k^2+k+2)(k-1)(k+2)^n-(k-2)(k+1)^2k^n}{2k(k+1)(k+2)} \nonumber\\
&-\frac{2}{(k+1)(k+2)}.
\end{align}
Substituting this result and $N_n=k^n$ into $H_1(\mathcal{S}(n,k))=\frac{\Theta_n}{2N_n}$ yields
\begin{align}\label{Siercoh13}
H_1(\mathcal{S}(n,k))=&\frac{(k^2+k+2)(k-1)(k+2)^n-4k}{4k^{n+1}(k+1)(k+2)}\nonumber\\
&-\frac{(k-2)(k+1)}{4k(k+2)}.
\end{align}
In the large limit of $n$,
\begin{equation}\label{Siercoh13a}
\lim_{n \to \infty}H_1(\mathcal{S}(n,k))= \frac{k^3+k-2}{4k(k+1)(k+2)}\left(\frac{k+2}{k}\right)^{n},
\end{equation}
which can be expressed in terms of  the network order $N_n$ as
\begin{equation}\label{Siercoh14}
\lim_{n \to \infty}H_1(\mathcal{S}(n,k))= \frac{k^3+k-2}{4k(k+1)(k+2)}N_n^{\frac{\log (k+2)}{\log k}-1}\,.
\end{equation}
This completes the proof.
\end{IEEEproof}

Theorem~\ref{siercoh1order} shows that for large Sierpi\'nski graphs $\mathcal{S}(n,k)$, the first-order coherence $H_1(\mathcal{S}(n,k))$ grows sublinearly with $N_n$. This is contrast to its counterpart of hierarchical graphs $\mathcal{H}(n,k)$, for which the first-order coherence increases logarithmically with $N_n$.

%{\color{red}
%Similarly to the hierarchical graphs, we can obtain the mean hitting time of Sierpi\'nski graphs $\mathcal{S}(n,k)$ as follows.
%\begin{theorem}\label{Siercohrw}
%For the Sierpi\'nski graphs $\mathcal{S}(n,k)$ with $N_n$ vertices, the mean hitting time $T(\mathcal{S}(n,k))$ scales with the number of vertices as
%\begin{equation}\label{Siercohrw1}
%T(\mathcal{S}(n,k))\thicksim \frac{k^3+k-2}{2(k+1)(k+2)}N_n^{\frac{\log (k+2)}{\log k}}\,.
%\end{equation}
%\end{theorem}
%}

\subsubsection{Second-order coherence}

We finally compute the second-order coherence in Sierpi\'nski graphs $\mathcal{S}(n,k)$.
\begin{theorem}\label{siercoh2order}
For the Sierpi\'nski graphs $S(n,k)$ with $N_n$ vertices, the second-order coherence of the system with dynamics defined in~\eqref{Con11} is
\begin{align}\label{Siercoh15a}
H_2(\mathcal{S}(n,k))=&-\frac{7k^2+13k+2}{2k^{n+1}(k+1)^2(k+2)^2(k+3)}-\nonumber\\
&\frac{(k-2)(k^3+4k^2+4k+2)}{2k^2(k+2)^2(k^2+3k+4)}+\nonumber\\
&\frac{(k-1)(k+2)^{n-2}(k^2+k+2)}{2k^{n+1}(k+1)^2}+\nonumber\\
&\frac{k^5+7k^4+16k^3+28k^2+26k+12}{2k^{n+2}(k+1)^2(k+3)(k^2+3k+4)}\cdot\nonumber\\
&(k-1)(k+2)^{2n-2}.
\end{align}
In large graphs ($n\rightarrow \infty$), the dominating term of $H_2(S(n,k))$ scales with network order $N_n$ as
\begin{equation}\label{Siercoh15b}
H_2(\mathcal{S}(n,k))\sim h(k) N_n^{\frac{2\log (k+2)}{\log k}-1},
\end{equation}
where $h(k)=\frac{(k^5+7k^4+16k^3+28k^2+26k+12)(k-1)}{2k^2(k+1)^2(k+2)^2(k+3)(k^2+3k+4)}$.
\end{theorem}
\vskip 0.25cm
\begin{IEEEproof}
Let $\Omega_n$ denote the sum of the reciprocals of square of all non-zero eigenvalues for Laplacian matrix of $\mathcal{S}(n,k)$. By definition,
\begin{equation}\label{Siercoh16}
\Omega_n=\sum_{i=2}^{N_n}\frac{1}{\left(\lambda_i^{(n)}\right)^2}=\sum_{\lambda_i^{(n)} \in \Phi_n^{(1)}}\frac{1}{\left(\lambda_i^{(n)}\right)^2}+\sum_{\widetilde{\lambda}_{i}^{(n)} \in \Phi_n^{(2)}}\frac{1}{\left(\widetilde{\lambda}_{i}^{(n)}\right)^2}.
\end{equation}
The first sum term on the rhs of~\eqref{Siercoh16} can be expressed as
\begin{equation}\label{Siercoh17}
\sum_{\lambda_i^{(n)} \in \Phi_n^{(1)}}\frac{1}{\left(\lambda_i^{(n)}\right)^2}=\frac{k-2}{2}\left(k^{n-3}+\frac{1}{k(k-2)}+\frac{k^{n-1}-1}{(k+2)^2} \right).
\end{equation}
%In addition,from~\eqref{Siercoh7} and~\eqref{Siercoh8}, one obtains
Using the following relation
\begin{equation}\label{Siercoh18}
\frac{1}{\left(\widetilde{\lambda}_{i,1}^{(n)}\right)^2}+\frac{1}{\left(\widetilde{\lambda}_{i,2}^{(n)}\right)^2}
%&=\frac{\left(\widetilde{\lambda}_{i,1}^{(n)}+\widetilde{\lambda}_{i,2}^{(n)}\right)^2-2\widetilde{\lambda}_{i,1}^{(n)}\cdot\widetilde{\lambda}_{i,2}^{(n)}}{\left(\widetilde{\lambda}_{i,1}^{(n)}\cdot\widetilde{\lambda}_{i,2}^{(n)}\right)^2}\nonumber\\
=\frac{(k+2)^2-2\lambda_{i}^{(n-1)}}{\left(\lambda_{i}^{(n-1)}\right)^2},
\end{equation}
 the second sum term in~\eqref{Siercoh16} can be evaluated as
\begin{align}\label{Siercoh19}
\sum_{\widetilde{\lambda}_{i}^{(n)} \in \Phi_n^{(2)}}\frac{1}{\left(\widetilde{\lambda}_{i}^{(n)}\right)^2}
&=\sum_{\lambda_{i}^{(n-1)} \in \Phi_{n-1}\setminus\{0\}}\frac{(k+2)^2-2\lambda_{i}^{(n-1)}}{\left(\lambda_{i}^{(n-1)}\right)^2}\nonumber\\
&=(k+2)^2\Omega_{n-1}-2\Theta_{n-1}.
\end{align}
Then, we have the following recursive relation for $\Omega_{n}$:
\begin{align}\label{Siercoh20}
\Omega_n=&(k+2)^2\Omega_{n-1}-2\Theta_{n-1}+\frac{k-2}{2}\bigg(k^{n-3}+\frac{1}{k(k-2)}+\nonumber\\
&\frac{k^{n-1}-1}{(k+2)^2} \bigg).
\end{align}
Applying~\eqref{Siercoh12} and the initial value $\Omega_{1}=\frac{k-1}{k^2}$,~\eqref{Siercoh20} is solved to obtain the explicit expression for $\Omega_n$.
Plugging the expression for $\Omega_n$ into relation $H_2(\mathcal{S}(n,k))=\frac{\Omega_n}{2N_n}$ yields~\eqref{Siercoh15a}.  %the exact formula for the second-order coherence in Sierpi\'nski graphs $\mathcal{S}(n,k)$ is obtained to be
%\begin{align}\label{Siercoh21}
%H_2(\mathcal{S}(n,k))=&-\frac{7k^2+13k+2}{2k^{n+1}(k+1)^2(k+2)^2(k+3)}-\nonumber\\
%&\frac{(k-2)(k^3+4k^2+4k+2)}{2k^2(k+2)^2(k^2+3k+4)}+\nonumber\\
%&\frac{(k-1)(k+2)^{n-2}(k^2+k+2)}{2k^{n+1}(k+1)^2}+\nonumber\\
%&\frac{k^5+7k^4+16k^3+28k^2+26k+12}{2k^{n+2}(k+1)^2(k+3)(k^2+3k+4)}\cdot\nonumber\\
%&(k-1)(k+2)^{2n-2}.
%\end{align}
For large $n$, the leading term of $H_2(\mathcal{S}(n,k))$ can be represented in terms of network order $N_n$ as
\begin{equation}\label{Siercoh22}
H_2(\mathcal{S}(n,k)) \sim h(k) N_n^{\frac{2\log (k+2)}{\log k}-1},
\end{equation}
where the factor $h(k)$ is a function of $k$:
\begin{equation}\label{Siercoh23}
h(k)=\frac{(k^5+7k^4+16k^3+28k^2+26k+12)(k-1)}{2k^2(k+1)^2(k+2)^2(k+3)(k^2+3k+4)}.
\end{equation}
This completes the proof.
\end{IEEEproof}

Theorem~\ref{siercoh2order}  indicates that the asymptotic behavior for the second-order coherence in the Sierpi\'nski graphs $\mathcal{S}(n,k)$ grows superlinearly with the network order $N_n$, and is considerably larger than that associated with the hierarchical graphs $\mathcal{H}(n,k)$.

\subsubsection{Analysis}

In the above, we have demonstrated that the behaviors of related quantities for consensus algorithms in hierarchical graphs  $\mathcal{H}(n,k)$ and Sierpi\'nski graphs  $\mathcal{S}(n,k)$ are strongly different from each other. The second smallest eigenvalue of  $\mathcal{H}(n,k)$ is considerably larger than that of  $\mathcal{S}(n,k)$, with their ratio being $\left (\frac{k+2}{k}\right)^{n-1}$ and increasing exponentially with $n$. The largest eigenvalue of  $\mathcal{H}(n,k)$ is remarkably greater than that of  $\mathcal{S}(n,k)$, with their ratio being $\frac{(k-1)n}{k+2} $ and growing linearly with $n$. In addition, the scalings of both the first-order and the second-order coherence in $\mathcal{H}(n,k)$ are smaller than those corresponding to $\mathcal{S}(n,k)$.

%{\color{blue}
%Because  both  hierarchical graphs  $\mathcal{H}(n,k)$ and Sierpi\'nski graphs  $\mathcal{S}(n,k)$  are self-similar, and have the same number of vertices and edges, but  differ in some structural aspects, such as degree distribution, average distance, and fractality, we argue that the difference for consensus algorithms  in these two graphs, lies in, at least partially, their  structural  discrepancy.
%}

Because  both  hierarchical graphs  $\mathcal{H}(n,k)$ and Sierpi\'nski graphs  $\mathcal{S}(n,k)$  are self-similar, and have the same number of vertices and edges, but  differ in some structural aspects, such as degree distribution, average distance, and fractality, we argue that the difference for consensus algorithms  in these two graphs, lies in, at least partially, their  structural  discrepancy. For example, hierarchical graphs exhibit the small-world effect, while Sierpi\'nski graph are ``large-world". The small-world structure can drastically reduce  communication time between different vertices and speed up  information diffusion in a network. Then, it is not difficult to understand that the speed of convergence of noiseless consensus algorithms in hierarchical graphs is faster than in Sierpi\'nski graphs, and the coherence for both first-order and second-order noise consensus problems is lower in hierarchical graphs than in Sierpi\'nski graphs.

%The distinction of the behaviors for consensus algorithms in hierarchical graphs  and Sierpi\'nski graphs lies in the structural difference between them.

%Since these quantities of consensus problems are closely related to the network topology, we demonstrate that these distinctions highlight the structural difference between these two graphs. By the construction of the graphs, the Sierpi\'nski graph are "large-world" and almost all of its vertices (except the $k$ extreme vertices), have the same degree and thus play a similar role in the dynamics process. Quite different from the Sierpi\'nski graphs, the hierarchical graphs exhibit small-world characteristics and heterogeneous degree distribution, in which there exist some hub vertices (i.e. high-degree vertices). These structure properties can drastically reduce the communication time between different vertices and speed up the information diffusion in the network. That is why both the communication delay and the coherence is lower in the hierarchical graphs than those in the Sierpi\'nski graph. These differences of dynamical properties indicate that heterogeneous degree distribution and small-world topologies have a strong effect on the robustness of noisy consensus algorithms.

\section{Conclusion}

The self-similarity property is ubiquitous in real-world and man-made systems. In this paper, we studied consensus problems in two iteratively growing self-similar networks, the hierarchical graphs and the Sierpi\'nski graphs. Both of these two networks have the same order and size at any iteration, but exhibit quite different topological properties. We studied in detail several important quantities of consensus problems in these two networks, including convergence speed, delay robustness, and coherence for first-order and second-order dynamics. We showed that the consensus problem can be solved faster in hierarchical graphs than in Sierpi\'nski graphs. In contrast, the hierarchical graphs can tolerate smaller communication delay than Sierpi\'nski graphs.

For the  first-order and second-order noisy consensus algorithms, the asymptotic behaviors of network coherence also scale differently in the two graphs. For the first-order noisy consensus algorithm, the network coherence grows logarithmically with the number $N$ of vertices in hierarchical graphs but sublinearly with network order $N$ in Sierpi\'nski graphs. For the second-order noisy consensus algorithm, the coherence grows linearly with $N$ in hierarchical graphs, but superlinearly with $N$ in Sierpi\'nski graphs.

We demonstrated that the structure difference of the  two self-similar networks  is responsible for  the observed distinct performance of the studied consensus algorithms defined on them.

\ifCLASSOPTIONcaptionsoff
  \newpage
\fi

\begin{IEEEbiography}[{\includegraphics[width=1in,height=1.25in,clip,keepaspectratio]{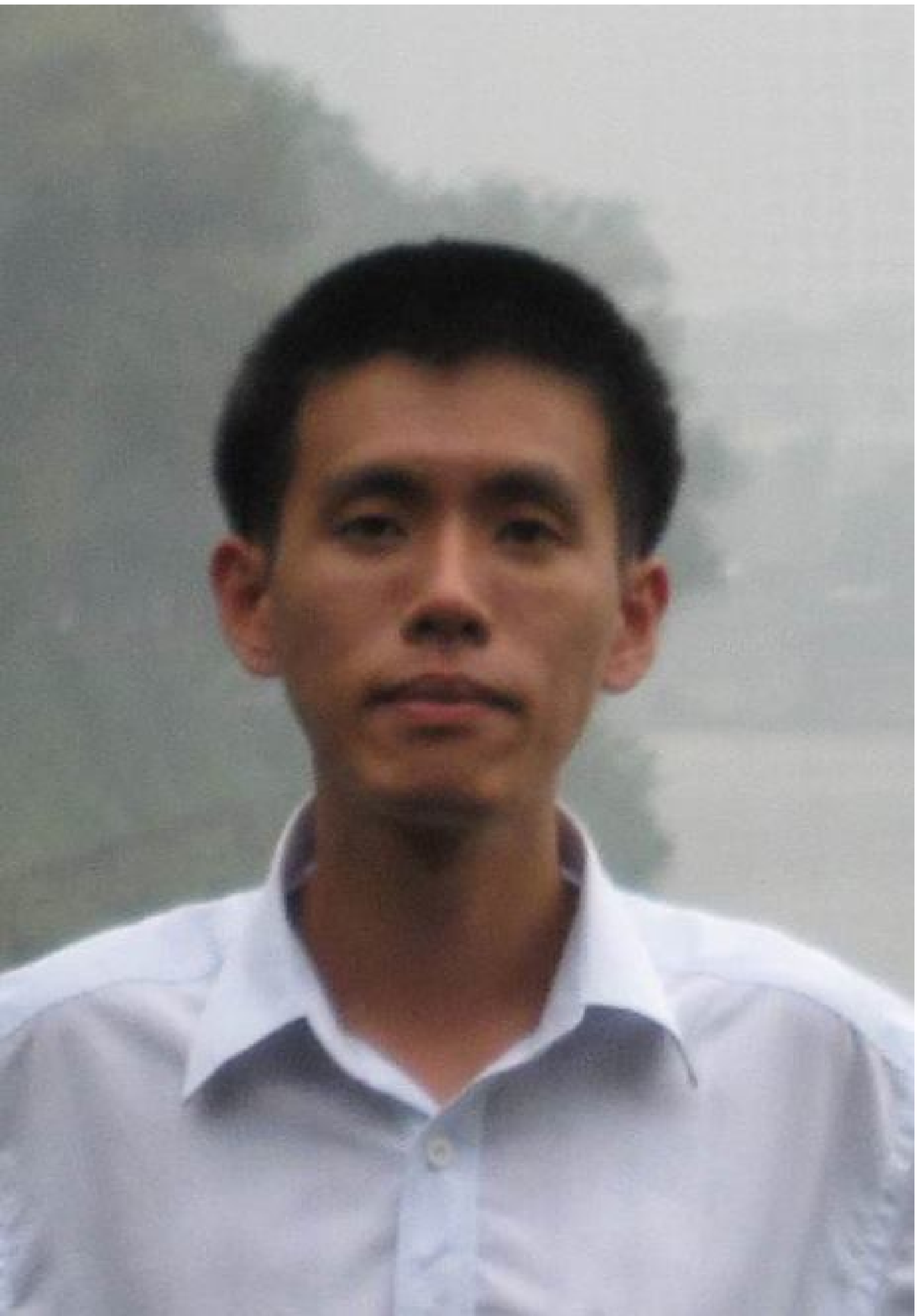}}]{Yi Qi}
received the B.Sc. degree in mathematics and the M.S. degree in computer science both from Fudan University, China, in 2007 and 2010, respectively. He worked as a researcher at the IBM China Research Laboratory from 2010 to 2013. He is currently working toward the Ph.D. degree in school of Computer Science, Fudan University. His research interests include network science, particularly in structural and dynamical properties of complex systems and networks.
\end{IEEEbiography}

\begin{IEEEbiography}[{\includegraphics[width=1in,height=1.25in,clip,keepaspectratio]{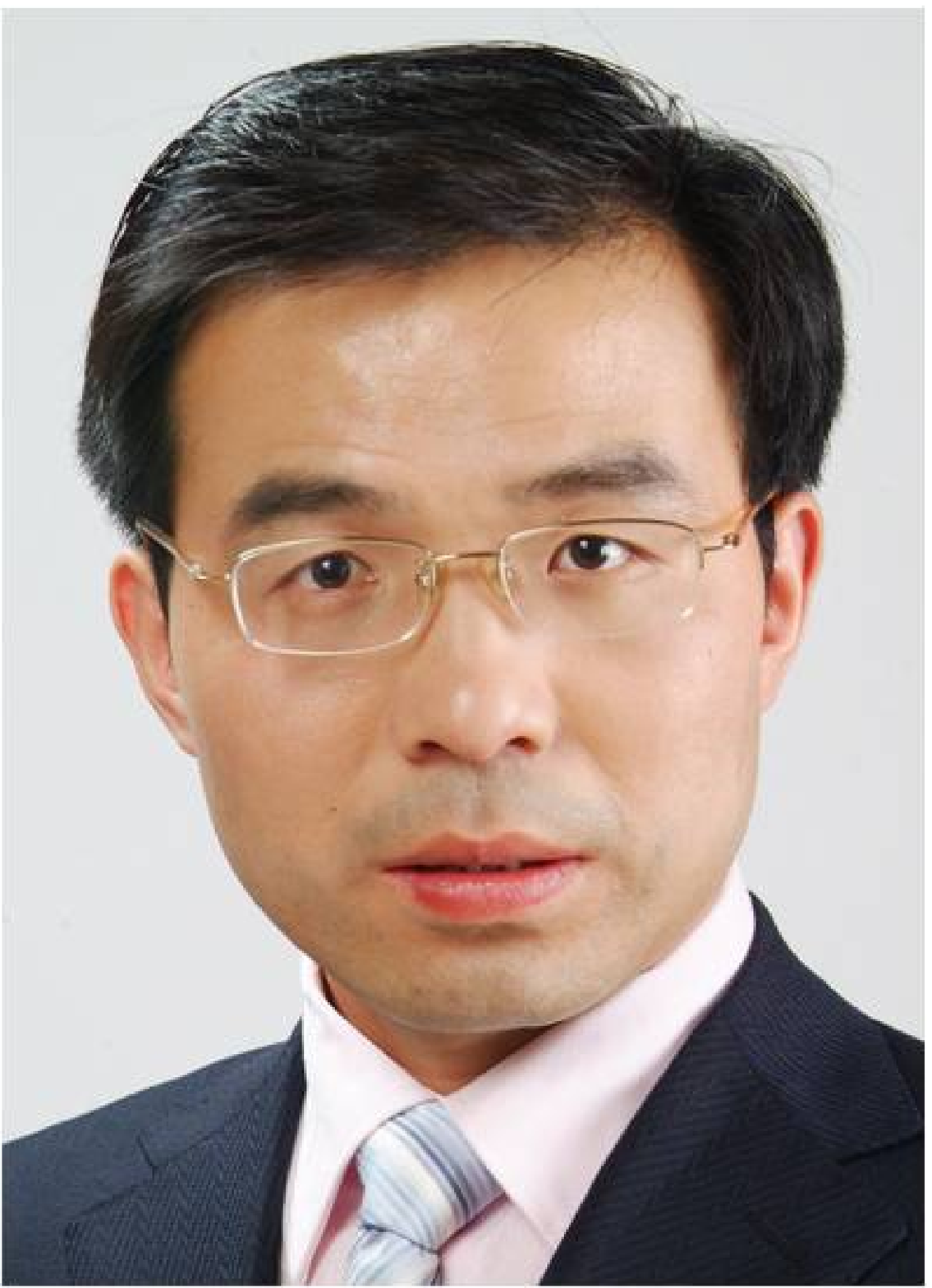}}]{Zhongzhi Zhang}
received the B.Sc. degree in Applied Mathematics from Anhui University, Hefei, China in 1997 and Ph.D. degree in Management Science and Engineering from Dalian University of Technology, Dalian, China, in 2006. From July 2006 to June 2008, he was a Post-Doctorate Research Fellow in Fudan University. Currently he is an Associate Professor at the School of Computer Science, Fudan University, China. Dr. Zhang's research interests include structural and dynamical properties of complex networks. He has published more than 100 papers in international journals since 2006 in the field of network modeling and dynamics, receiving over 2000 citations with H index 27 according to the Thomson Reuters ISI Web of Science.  Dr. Zhang is a Committee Member of the Chinese Society of Complexity Science. He received the Excellent Doctoral Dissertation Award of Liaoning Province, China in 2007, the Excellent Post-Doctor Award of Fudan University in 2008, and the Shanghai Natural Science Award (3rd class) in 2013.
\end{IEEEbiography}

\begin{IEEEbiography}[{\includegraphics[width=1in,height=1.25in,clip,keepaspectratio]{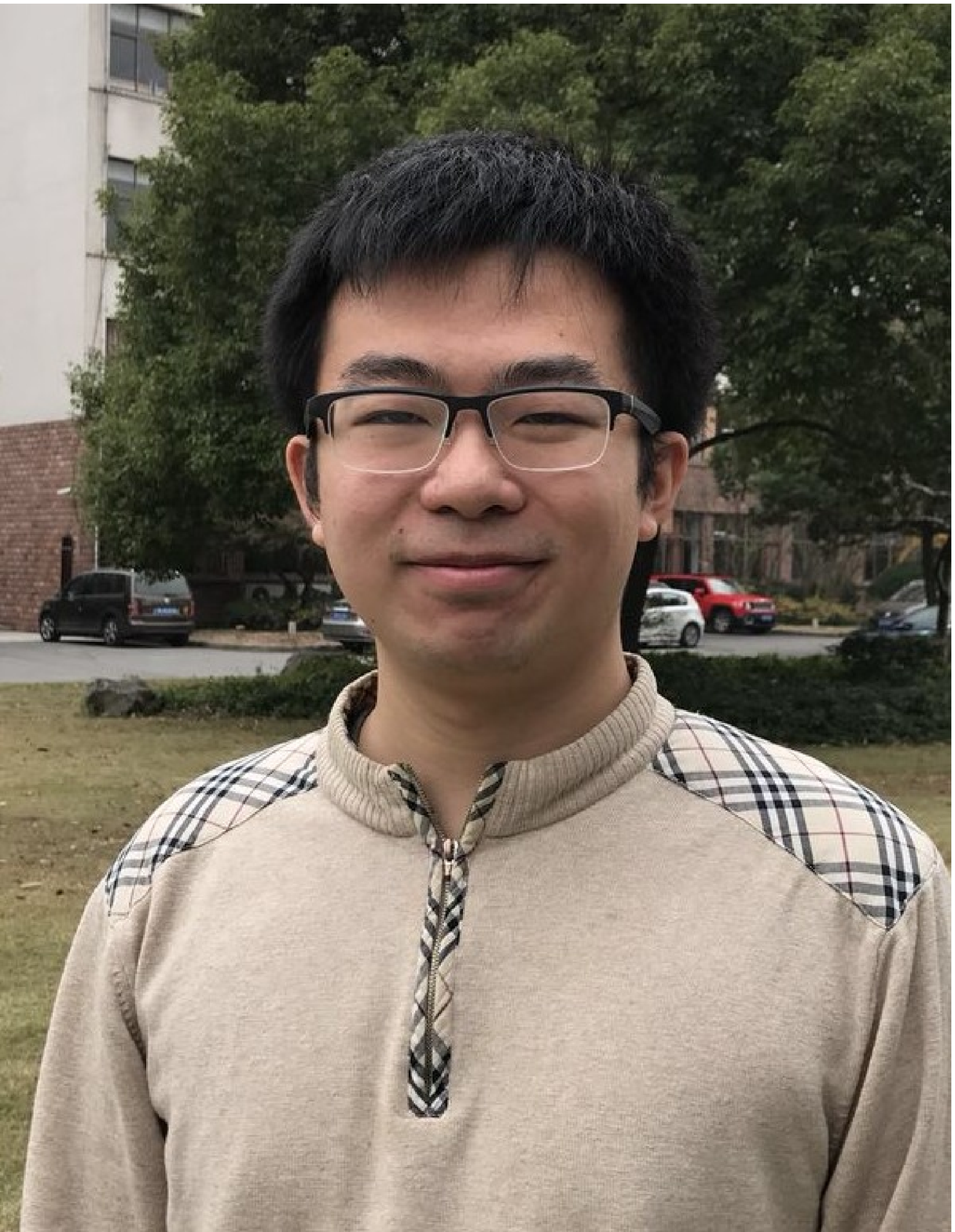}}]{Yuhao Yi}
received the B.Sc. degree in Computer Science from Fudan University, Shanghai, China, in 2013, where he is currently pursuing
the Ph.D. degree. His research interests include network science and distributed control. His current work centers on analyzing and optimizing the performance of averaging consensus dynamics in large-scale networks.
\end{IEEEbiography}

\begin{IEEEbiography}[{\includegraphics[width=1in,height=1.25in,clip,keepaspectratio]{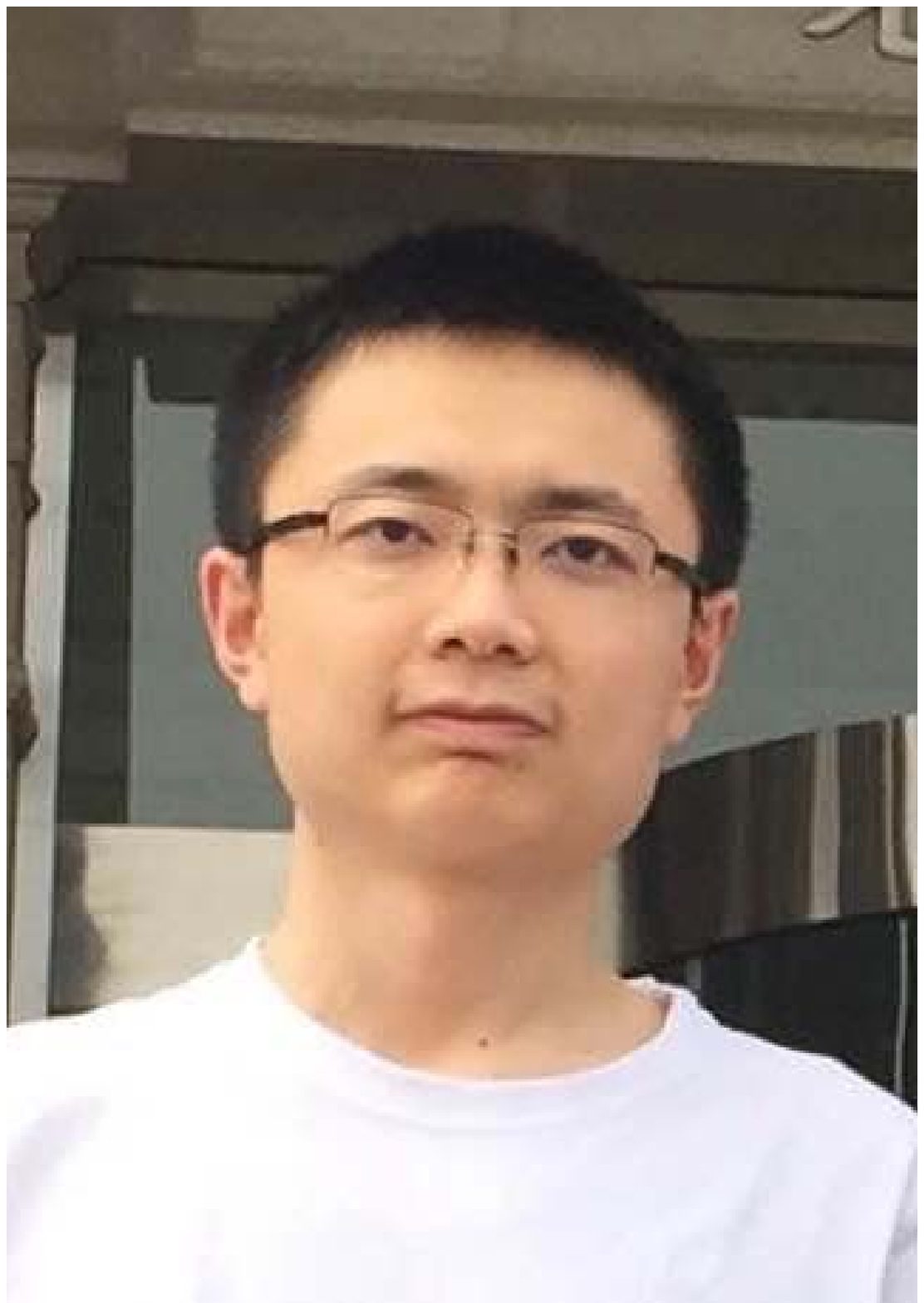}}]{Huan Li}
received the B.S. degree in School of Computer Science, Fudan University, Shanghai, China, in 2016, where he is currently pursuing the Master's degree.
His research interests include graph algorithms, social networks, and network science.
\end{IEEEbiography}


%\vfill



\begin{thebibliography}{10}
\providecommand{\url}[1]{#1}
\csname url@samestyle\endcsname
\providecommand{\newblock}{\relax}
\providecommand{\bibinfo}[2]{#2}
\providecommand{\BIBentrySTDinterwordspacing}{\spaceskip=0pt\relax}
\providecommand{\BIBentryALTinterwordstretchfactor}{4}
\providecommand{\BIBentryALTinterwordspacing}{\spaceskip=\fontdimen2\font plus
\BIBentryALTinterwordstretchfactor\fontdimen3\font minus
  \fontdimen4\font\relax}
\providecommand{\BIBforeignlanguage}[2]{{%
\expandafter\ifx\csname l@#1\endcsname\relax
\typeout{** WARNING: IEEEtran.bst: No hyphenation pattern has been}%
\typeout{** loaded for the language `#1'. Using the pattern for}%
\typeout{** the default language instead.}%
\else
\language=\csname l@#1\endcsname
\fi
#2}}
\providecommand{\BIBdecl}{\relax}
\BIBdecl

\bibitem{De74}
M.~H. DeGroot, ``Reaching a consensus,'' \emph{J. Am. Statist. Assoc.},
  vol.~69, no. 345, pp. 118--121, 1974.

\bibitem{Ly97}
N.~A. Lynch, \emph{Distributed algorithms}.\hskip 1em plus 0.5em minus
  0.4em\relax San Francisco, CA: Morgan Kaufmann, 1996.

\bibitem{CrNoGuToRi03}
G.~Korniss, M.~Novotny, H.~Guclu, Z.~Toroczkai, and P.~A. Rikvold,
  ``Suppressing roughness of virtual times in parallel discrete-event
  simulations,'' \emph{Science}, vol. 299, no. 5607, pp. 677--679, 2003.

\bibitem{Cy89}
G.~Cybenko, ``Dynamic load balancing for distributed memory multiprocessors,''
  \emph{J. Parallel Distr. Comput.}, vol.~7, no.~2, pp. 279--301, 1989.

\bibitem{MuGhSc98}
S.~Muthukrishnan, B.~Ghosh, and M.~H. Schultz, ``First-and second-order
  diffusive methods for rapid, coarse, distributed load balancing,''
  \emph{Theor. Comput. Syst.}, vol.~31, no.~4, pp. 331--354, 1998.

\bibitem{DiFrMo99}
R.~Diekmann, A.~Frommer, and B.~Monien, ``Efficient schemes for nearest
  neighbor load balancing,'' \emph{Parallel Comput.}, vol.~25, no.~7, pp.
  789--812, 1999.

\bibitem{LiRu06}
Q.~Li and D.~Rus, ``Global clock synchronization in sensor networks,''
  \emph{IEEE Trans. Comput.}, vol.~55, no.~2, pp. 214--226, 2006.

\bibitem{YuChWaYa09}
W.~Yu, G.~Chen, Z.~Wang, and W.~Yang, ``Distributed consensus filtering in
  sensor networks,'' \emph{IEEE Trans. Syst., Man, Cybern. B, Cybern.},
  vol.~39, no.~6, pp. 1568--1577, 2009.

\bibitem{ZhChLiYaGu13}
S.~Zhu, C.~Chen, W.~Li, B.~Yang, and X.~Guan, ``Distributed optimal consensus
  filter for target tracking in heterogeneous sensor networks,'' \emph{IEEE
  Trans. Cybern.}, vol.~43, no.~6, pp. 1963--1976, 2013.

\bibitem{Sa06}
R.~Olfati-Saber, ``Flocking for multi-agent dynamic systems: Algorithms and
  theory,'' \emph{IEEE Trans. Autom. Control}, vol.~51, no.~3, pp. 401--420,
  Mar. 2006.

\bibitem{DiKy07}
D.~V. Dimarogonas and K.~J. Kyriakopoulos, ``On the rendezvous problem for
  multiple nonholonomic agents,'' \emph{IEEE Trans. Autom. Control}, vol.~52,
  no.~5, pp. 916--922, May 2007.

\bibitem{FaMu04}
J.~A. Fax and R.~M. Murray, ``Information flow and cooperative control of
  vehicle formations,'' \emph{IEEE Trans. Autom. Control}, vol.~49, no.~9, pp.
  1465--1476, Sep. 2004.

\bibitem{JaLiMo03}
A.~Jadbabaie, J.~Lin, and A.~S. Morse, ``Coordination of groups of mobile
  autonomous agents using nearest neighbor rules,'' \emph{IEEE Trans. Autom.
  Control}, vol.~48, no.~6, pp. 988--1001, Jun. 2003.

\bibitem{OlMu04}
R.~Olfati-Saber and R.~M. Murray, ``Consensus problems in networks of agents
  with switching topology and time-delays,'' \emph{IEEE Trans. Autom. Control},
  vol.~49, no.~9, pp. 1520--1533, Sep. 2004.

\bibitem{Mo05}
L.~Moreau, ``Stability of multiagent systems with time-dependent communication
  links,'' \emph{IEEE Trans. Autom. Control}, vol.~50, no.~2, pp. 169--182,
  Feb. 2005.

\bibitem{IlStSt14}
N.~Ilic, M.~S. Stankovic, and S.~S. Stankovic, ``Adaptive consensus-based
  distributed target tracking in sensor networks with limited sensing range,''
  \emph{IEEE Trans. Control Syst. Technol.}, vol.~22, no.~2, pp. 778--785,
  2014.

\bibitem{KiAh15}
B.-Y. Kim and H.-S. Ahn, ``Consensus-based coordination and control for
  building automation systems,'' \emph{IEEE Trans. Control Syst. Technol.},
  vol.~23, no.~1, pp. 364--371, 2015.

\bibitem{DaGh15}
K.~Das and D.~Ghose, ``Broadcast control mechanism for positional consensus in
  multiagent systems,'' \emph{IEEE Trans. Control Syst. Technol.}, vol.~23,
  no.~5, pp. 1807--1826, 2015.

\bibitem{MiSt90}
R.~E. Mirollo and S.~H. Strogatz, ``Synchronization of pulse-coupled biological
  oscillators,'' \emph{SIAM J. Appl. Math.}, vol.~50, no.~6, pp. 1645--1662,
  1990.

\bibitem{SaFaMu07}
R.~Olfati-Saber, J.~A. Fax, and R.~M. Murray, ``Consensus and cooperation in
  networked multi-agent systems,'' \emph{Proc. IEEE}, vol.~95, no.~1, pp.
  215--233, Jan. 2007.

\bibitem{WuTaCaZh16}
X.~Wu, Y.~Tang, J.~Cao, and W.~Zhang, ``Distributed consensus of stochastic
  delayed multi-agent systems under asynchronous switching,'' \emph{IEEE Trans.
  Cybern.}, vol.~46, no.~8, pp. 1817--1827, 2016.

\bibitem{Ol05}
R.~Olfati-Saber, ``Ultrafast consensus in small-world networks,'' in
  \emph{Proc. Amer. Control Conf.}\hskip 1em plus 0.5em minus 0.4em\relax IEEE,
  2005, pp. 2371--2378.

\bibitem{OlTs09}
A.~Olshevsky and J.~N. Tsitsiklis, ``Convergence speed in distributed consensus
  and averaging,'' \emph{SIAM J. Control Optim.}, vol.~48, no.~1, pp. 33--55,
  2009.

\bibitem{AyBa10}
T.~C. Aysal and K.~E. Barner, ``Convergence of consensus models with stochastic
  disturbances,'' \emph{IEEE Trans. Inf. Theory}, vol.~56, no.~8, pp.
  4101--4113, 2010.

\bibitem{XiWa08}
F.~Xiao and L.~Wang, ``Consensus protocols for discrete-time multi-agent
  systems with time-varying delays,'' \emph{Automatica}, vol.~44, no.~10, pp.
  2577--2582, 2008.

\bibitem{MuPaAl10}
U.~M{\"u}nz, A.~Papachristodoulou, and F.~Allg{\"o}wer, ``Delay robustness in
  consensus problems,'' \emph{Automatica}, vol.~46, no.~8, pp. 1252--1265,
  2010.

\bibitem{XiBoKi07}
L.~Xiao, S.~Boyd, and S.-J. Kim, ``Distributed average consensus with
  least-mean-square deviation,'' \emph{J. Parallel Dist. Comput.}, vol.~67,
  no.~1, pp. 33--46, Jan. 2007.

\bibitem{BaJoMiPa08}
B.~Bamieh, M.~Jovanovic~R, P.~Mitra, and S.~Patterson, ``Effect of topological
  dimension on rigidity of vehicle formations: Fundamental limitations of local
  feedback,'' in \emph{Proc. 47th {IEEE} Conf. Decision Control}, Dec. 2008,
  pp. 369--374.

\bibitem{YoScLe10}
G.~F. Young, L.~Scardovi, and N.~E. Leonard, ``Robustness of noisy consensus
  dynamics with directed communication,'' in \emph{Proc. Amer. Control Conf.},
  Jun. 2010, pp. 6312--6317.

\bibitem{PaBa10}
S.~Patterson and B.~Bamieh, ``Leader selection for optimal network coherence,''
  in \emph{Proc. 49th {IEEE} Conf. Decision Control}.\hskip 1em plus 0.5em
  minus 0.4em\relax IEEE, 2010, pp. 2692--2697.

\bibitem{BaJoMiPa12}
B.~Bamieh, M.~Jovanovic~R, P.~Mitra, and S.~Patterson, ``Coherence in
  large-scale networks: Dimension-dependent limitations of local feedback,''
  \emph{{IEEE} Trans. Autom. Control}, vol.~57, no.~9, pp. 2235--2249, Sep.
  2012.

\bibitem{PaBa14}
S.~Patterson and B.~Bamieh, ``Consensus and coherence in fractal networks,''
  \emph{IEEE Trans. Control Netw. Syst.}, vol.~1, no.~4, pp. 338--348, Sep.
  2014.

\bibitem{OlMu03}
R.~O. Saber and R.~M. Murray, ``Consensus protocols for networks of dynamic
  agents,'' in \emph{Proc. Amer. Control Conf.}\hskip 1em plus 0.5em minus
  0.4em\relax IEEE, 2003, pp. 951--956.

\bibitem{GrMeSu90}
R.~Grone, R.~Merris, and V.~S. Sunder, ``{The Laplacian spectrum of a graph},''
  \emph{SIAM J. Matrix Anal. Appl.}, vol.~11, no.~2, pp. 218--238, 1990.

\bibitem{WaSt98}
D.~J. Watts and S.~H. Strogatz, ``Collective dynamics of `small-world'
  networks,'' \emph{Nature}, vol. 393, no. 6684, pp. 440--442, {Jun.} 1998.

\bibitem{Ne03}
M.~E.~J. Newman, ``{The structure and function of complex networks},''
  \emph{{SIAM Rev.}}, vol.~{45}, no.~{2}, pp. {167--256}, {Jun.} {2003}.

\bibitem{BaCoDaFi09}
L.~Barri{\`e}re, F.~Comellas, C.~Dalf{\'o}, and M.~A. Fiol, ``The hierarchical
  product of graphs,'' \emph{Discrete Appl. Math.}, vol. 157, pp. 36--48, 2009.

\bibitem{KlMi97}
S.~Klav{\v{z}}ar and U.~Milutinovi{\'c}, ``Graphs {$S(n,k)$} and a variant of
  the {T}ower of {H}anoi problem,'' \emph{Czech. Math. J.}, vol.~47, no.~1, pp.
  95--104, 1997.

\bibitem{DeKr12}
S.~DeDeo and D.~C. Krakauer, ``Dynamics and processing in finite self-similar
  networks,'' \emph{J. R. Soc. Interface}, vol.~9, no.~74, pp. 2131--2144,
  2012.

\bibitem{Fi7303}
M.~Fiedler, ``Algebraic connectivity of graphs,'' \emph{Czechoslovak Math. J.},
  vol.~23, no.~23, pp. 298--305, 1973.

\bibitem{HuKoSz10}
D.~Hunt, G.~Korniss, and B.~K. Szymanski, ``{Network synchronization in a noisy
  environment with time delays: Fundamental limits and trade-offs},''
  \emph{Phys. Rev. Lett.}, vol. 105, no.~6, p. 068701, 2010.

\bibitem{PaBa11}
S.~Patterson and B.~Bamieh, ``Network coherence in fractal graphs,'' in
  \emph{Proc. 50th {IEEE} Conf. Decision Control}, Dec. 2011, pp. 6445--6450.

\bibitem{DeMeRoSaVa18}
P.~De~Meo, F.~Messina, D.~Rosaci, G.~M. Sarn{\'e}, and A.~V. Vasilakos,
  ``Estimating graph robustness through the {R}andic index,'' \emph{IEEE Trans.
  Cybern.}, vol.~**, no.~*, pp. ****--****, 2018.

\bibitem{KlRa93}
D.~J. Klein and M.~Randi{\'c}, ``Resistance distance,'' \emph{J. Math. Chem.},
  vol.~12, no.~1, pp. 81--95, 1993.

\bibitem{LiLiCa16}
Q.~Liu, J.-B. Liu, and J.~Cao, ``{The Laplacian polynomial and Kirchhoff index
  of graphs based on R-graphs},'' \emph{Neurocomputing}, vol. 177, pp.
  441--446, 2016.

\bibitem{Te91}
P.~Tetali, ``Random walks and the effective resistance of networks,'' \emph{J.
  Theor. Probab.}, vol.~4, no.~1, pp. 101--109, 1991.

\bibitem{ChRaRuSm89}
A.~K. Chandra, P.~Raghavan, W.~L. Ruzzo, and R.~Smolensky, ``The electrical
  resistance of a graph captures its commute and cover times,'' in \emph{Proc.
  21st {ACM} Symp. Theory of Computing}, 1989, pp. 574--586.

\bibitem{MaEl15}
X.~Ma and N.~Elia, ``Mean square performance and robust yet fragile nature of
  torus networked average consensus,'' \emph{IEEE Trans. Control Netw. Syst.},
  vol.~2, no.~3, pp. 216--225, 2015.

\bibitem{YiZhLiCh15}
Y.~Yi, Z.~Zhang, Y.~Lin, and G.~Chen, ``Small-world topology can significantly
  improve the performance of noisy consensus in a complex network,''
  \emph{Comput. J.}, vol.~58, no.~12, pp. 3242--3254, 2015.

\bibitem{ZhCo11}
Z.~Zhang and F.~Comellas, ``Farey graphs as models for complex networks,''
  \emph{Theor. Comput. Sci.}, vol. 412, no.~8, pp. 865--875, Mar. 2011.

\bibitem{YiZhShCh17}
Y.~Yi, Z.~Zhang, L.~Shan, and G.~Chen, ``Robustness of first-and second-order
  consensus algorithms for a noisy scale-free small-world {K}och network,''
  \emph{IEEE Trans. Control Syst. Technol.}, vol.~25, no.~1, pp. 342--350,
  2017.

\bibitem{SuShLyDo15}
T.~Summers, I.~Shames, J.~Lygeros, and F.~D{\"o}rfler, ``Topology design for
  optimal network coherence,'' in \emph{Proc. Eur. Control Conf.}, 2015, pp.
  575--580.

\bibitem{CaJaSo01}
R.~F.~i. Cancho, C.~Janssen, and R.~V. Sol\'e, ``Topology of technology graphs:
  Small world patterns in electronic circuits,'' \emph{Phys. Rev. E}, vol.~64,
  p. 046119, 2001.

\bibitem{AmScBaSt00}
L.~A.~N. Amaral, A.~Scala, M.~Barthelemy, and H.~E. Stanley, ``Classes of
  small-world networks,'' \emph{Proc. Natl. Acad. Sci.}, vol.~97, no.~21, pp.
  11\,149--11\,152, 2000.

\bibitem{BaAl99}
A.-L. Barab{\'a}si and R.~Albert, ``Emergence of scaling in random networks,''
  \emph{Science}, vol. 286, no. 5439, pp. 509--512, 1999.

\bibitem{GoMc78}
C.-D. Godsil and B.-D. McKay, ``A new graph product and its spectrum,''
  \emph{Bull. Austral. Math. Soc.}, vol.~18, no.~01, pp. 21--28, 1978.

\bibitem{liDoQiZh15}
H.~Liu, M.~Dolgushev, Y.~Qi, and Z.~Zhang, ``Laplacian spectra of a class of
  small-world networks and their applications,'' \emph{Sci. Rep.}, vol.~5, p.
  9024, 2015.

\bibitem{Kr05}
B.~Krön, ``Growth of self-similar graphs,'' \emph{J. Graph Theory}, vol.~45,
  no.~3, pp. 224--239, 2002.

\bibitem{SoHaMa05}
C.~Song, S.~Havlin, and H.~Makse, ``Self-similarity of complex networks,''
  \emph{Nature}, vol. 433, no. 7024, pp. 392--395, 2005.

\bibitem{ZhWuLiCo16}
Z.~Zhang, S.~Wu, M.~Li, and F.~Comellas, ``{The number and degree distribution
  of spanning trees in the Tower of Hanoi graph},'' \emph{Theoret. Comput.
  Sci.}, vol. 609, pp. 443--455, 2016.

\bibitem{JiLiZh17}
Y.~Jin, H.~Li, and Z.~Zhang, ``{Maximum matchings and minimum dominating sets
  in Apollonian networks and extended Tower of Hanoi graphs},'' \emph{Theoret.
  Comput. Sci.}, vol. 703, pp. 37--54, 2017.

\bibitem{Te90}
A.~Telcs, ``{Spectra of graphs and fractal dimensions I},'' \emph{Probab.
  Theory Rel. Fields}, vol.~85, pp. 489--497, 1990.

\bibitem{MaPe97}
U.~M.~B. Marconi and A.~Petri, ``{Time dependent Ginzburg-Landau model in the
  absence of translational invariance. Non-conserved order parameter domain
  growth},'' \emph{J. Phys. A}, vol.~30, no.~4, p. 1069, 1997.

\bibitem{KlMo05}
S.~Klav{\v{z}}ar and B.~Mohar, ``Crossing numbers of {S}ierpi{\'n}ski-like
  graphs,'' \emph{J. Graph Theory}, vol.~50, pp. 186--198, 2005.

\bibitem{HiKlMiPeSt13}
A.~M. Hinz, S.~Klav{\v{z}}ar, U.~Milutinovi{\'c}, and C.~Petr, \emph{The Tower
  of {H}anoi-- Myths and Maths}.\hskip 1em plus 0.5em minus 0.4em\relax
  Springer, 2013.

\bibitem{HiklSa17}
A.~M. Hinz, S.~Klav{\v{z}}ar, and S.~S. Zemlji{\v{c}}, ``{A survey and
  classification of Sierpi{\'n}ski-type graphs},'' \emph{Discrete Appl. Math.},
  vol. 217, pp. 565--600, 2017.

\bibitem{FaImSa08}
M.~H. Farahabady, N.~Imani, and H.~Sarbazi-Azad, ``{Some topological and
  combinatorial properties of WK-recursive mesh and WK-pyramid interconnection
  networks},'' \emph{J. Syst. Archit.}, vol.~54, no.~10, pp. 967--976, 2008.

\bibitem{WaJu15}
Y.-C. Wang and J.~S.-T. Juan, ``Hamiltonicity of the basic {WK}-recursive
  pyramid with and without faulty nodes,'' \emph{Theoret. Comput. Sci.}, vol.
  562, pp. 542--556, 2015.

\bibitem{FaWaHu07}
J.~F. Fang, Y.~R. Wang, and H.~L. Huang, ``The $m$-pancycle-connectivity of a
  {WK}-{R}ecursive network,'' \emph{Inform. Sci.}, vol. 177, no.~24, pp.
  5611--5619, 2007.

\bibitem{FaLiChNg07}
J.~F. Fang, W.~Y. Liang, H.~R. Chen, and K.~L. Ng, ``Novel broadcasting
  algorithm of the complete recursive network,'' \emph{Informatica}, vol.~31,
  no.~1, pp. 131--136, 2007.

\bibitem{Fu05}
J.~S. Fu, ``Hamiltonicity of the {WK}-{R}ecursive network with and without
  faulty nodes,'' \emph{IEEE. Trans. Parallel Distr.}, vol.~16, no.~9, pp.
  853--865, 2005.

\bibitem{BlJu02}
A.~Blumen and A.~Jurjiu, ``Multifractal spectra and the relaxation of model
  polymer networks,'' \emph{J. Chem. Phys.}, vol. 116, no.~6, pp. 2636--2641,
  2002.

\end{thebibliography}
\end{document}